\documentclass[aps,prl,twocolumn,superscriptaddress,showpacs,longbibliography]{revtex4-1}
\usepackage[colorlinks=true,citecolor=blue,linkcolor=blue,breaklinks=true]{hyperref}
\usepackage{amssymb}
\usepackage{graphicx}
\usepackage{amsmath}
\usepackage{mathtools}
\usepackage[export]{adjustbox}
\usepackage{epsfig}
\usepackage{times}
\usepackage{color}
\usepackage{subfigure}
\usepackage{setspace}
\usepackage{bm}
\usepackage{calc}
\usepackage{natbib}
\begin{document}

\definecolor{ao}{rgb}{0.0, 0.5, 0.0}
\newcommand{\acc}[1]{\textcolor{blue}{\textsf{[AC: #1]}}}
\newcommand{\ac}[1]{\textcolor{blue}{\textsf{ #1}}}
\newcommand{\acd}[1]{\textcolor{ao}{\textsf{ #1}}}
\newcommand{\fm}[1]{\textcolor{black}{#1}}
\newcommand {\ba} {\ensuremath{b^\dagger}}
\newcommand {\Ma} {\ensuremath{M^\dagger}}
\newcommand {\psia} {\ensuremath{\psi^\dagger}}
\newcommand {\psita} {\ensuremath{\tilde{\psi}^\dagger}}
\newcommand{\lp} {\ensuremath{{\lambda '}}}
\newcommand{\A} {\ensuremath{{\bf A}}}
\newcommand{\Q} {\ensuremath{{\bf Q}}}
\newcommand{\kk} {\ensuremath{{\bf k}}}
\newcommand{\qq} {\ensuremath{{\bf q}}}
\newcommand{\kp} {\ensuremath{{\bf k'}}}
\newcommand{\rr} {\ensuremath{{\bf r}}}
\newcommand{\rp} {\ensuremath{{\bf r'}}}
\newcommand {\ep} {\ensuremath{\epsilon}}
\newcommand{\nbr} {\ensuremath{\langle ij \rangle}}
\newcommand {\no} {\nonumber}
\newcommand{\up} {\ensuremath{\uparrow}}
\newcommand{\dn} {\ensuremath{\downarrow}}
\newcommand{\rcol} {\textcolor{red}}

\newcommand{\fpc}[1]{{\color{red}[  \texttt{#1}]}}
\newcommand{\fpm}[1]{{\color{blue}  #1}} 


\begin{abstract}
Recently, the possibility of inducing
superconductivity for electrons in two dimensional materials has been
proposed via cavity-mediated pairing. The cavity-mediated electron-electron
interactions are long range, which has two main effects: firstly,
within the standard BCS-type pairing mediated by adiabatic photons, the superconducting critical temperature
depends polynomially on the coupling strength, instead of the
exponential dependence characterizing the phonon-mediated pairing;
secondly, as we show here, the effect of photon fluctuations is significantly
enhanced. These mediate novel non-BCS-type pairing processes, via non-adiabatic photons,
which are not sensitive to the electron occupation but rather to the
electron dispersion and lifetime at the Fermi
surface. 
Therefore, while the leading temperature dependence of BCS pairing comes from the smoothening of the Fermi-Dirac distribution, the temperature dependence of the fluctuation-induced pairing comes from the electron lifetime.
For realistic parameters, also including cavity loss, 
this results into a critical temperature which can be more than one
order of magnitude larger than the BCS prediction.
Moreover, a finite average number photons (as can be
achieved by incoherently pumping the cavity) adds to the
fluctuations and leads to a further enhancement of the critical temperature. 

 \end{abstract}
\title{Long-range photon fluctuations enhance photon-mediated electron pairing and superconductivity}
\author{Ahana Chakraborty}
\author{Francesco Piazza}
 \affiliation{Max Planck Institute for the Physics of Complex Systems, N\"othnitzer Str. 38, 01187, Dresden, Germany.}

\date{\today}

\maketitle

{\it Introduction. }
The development of experimental solid state platforms coupling
electrons with the
quantum light of optical cavities
\cite{Paravicini-Bagliani2019,Thomas2019} offers exciting prospects for the exploration of novel types of
collective phenomena, which can arise due to the peculiar nature of the
cavity-mediated interactions and their interplay with 
electronic correlations. Several scenarios have been
theoretically investigated, 
with relevance for solid-state materials \cite{Sentef2018,Curtis2019,Mazza2019,Kiffner2019,Schlawin2019,Allocca2019,Gao2020,Ashida2020,Sentef2020} and ultracold fermionic
atoms \cite{Piazza2014Umklapp,Keeling2014Fermionic,Chen2014Superradiance,Piazza2014Quantum,Pan2015Topological,Zheng2016Superradiance,Kollath2016Ultracold,Mivehvar2017Superradiant,Colella2018Quantum,Mivehvar2019Cavity,Schlawin2019a}.

One particular direction which is receiving considerable attention is the possibility to induce electronic
superconductivity via photon-mediated pairing
\cite{Schlawin2019,Gao2020}.
The critical temperature has been predicted to follow a
  non-exponential dependence on the light-matter coupling strength, which can be explained within
the usual BCS paradigm as due to the long-range character of the
cavity-mediated interactions i.e. the fact that the photons 
transfer a well defined momentum.

In this letter, we show that the long-range
nature of the cavity-mediated interactions can have an even more dramatic
influence on superconductivity, which is to introduce a novel, non-BCS-type of
pairing mediated by on-shell, non-adiabatic photon fluctuations. This is different from
BCS pairing, which involves instead the emission/absorption of
adiabatic, off-shell photons, and depends on the thermal occupation
of electrons. On the other hand, the fluctuation-mediated pairing is only
directly affected by the electron dispersion and lifetime near the Fermi
surface.
Therefore, temperature hinders this non-BCS-type of pairing by decreasing the
electron lifetime, while it affects the BCS pairing mainly by smoothening the Fermi-Dirac distribution.

This new fluctuation-mediated pairing can enhance superconductivity significantly.
Considering a Fermi-liquid-type electron lifetime set by the screened Coulomb
interaction, the critical temperature can be more than one
order of magnitude larger than the BCS prediction, using realistic
parameters for the terahertz cavities considered in Ref.\cite{Gao2020} and including photon
loss.

At the typical transition temperatures, optical cavities are unoccupied on average,
such that only vacuum fluctuations or loss-induced noise contribute to
the non-BCS-type pairing. A finite average number of photons can be
achieved by pumping the cavity, which then further amplifies this type
of pairing. In this way, the critical temperature can be
enhanced at least until the Fermi-liquid picture holds, approaching thus
the Fermi-temperature (slightly above 10 K for the materials considered in Ref.\cite{Gao2020}), even at
moderate electron-photon coupling.

{\it Model.} We consider a 2D electron system coupled to a terahertz cavity. The phenomenology discussed here relies on two properties of the cavity field. The first and most important feature is that the cavity-mediated interaction
  is long ranged. We model this by restricting the momentum
  transferred by the photon to a fixed vector $\vec{q}_0$. This simplest
  model corresponds to the case of a split-ring cavity 
  \cite{Gao2020}, sustaining a standing-wave mode of frequency
$\omega_c$, and momentum $\pm q_0 \hat{x}$, with $q_0=\omega_c
\sqrt{\epsilon_r}/c$. Here, $\epsilon_r$ is the relative permittivity
of the 2D material and $c$ is speed of light in vacuum.
We shall show that our results remain essentially unchanged in the
  limit $q_0\to 0$ as well as by allowing for a broadening of the photon momentum,
provided the corresponding frequency width is smaller than
  the electron lifetime at the Fermi surface (FS). This extends the applicability of our results
  to the case of the planar microcavity geometry considered in Ref.\cite{Schlawin2019}, where
the set of propagating transverse modes is distributed around $q=0$
with a narrow width set by $\omega_c/c$, where now $\omega_c$ is the
frequency of the purely longitudinal mode. The second feature we consider here is that the cavity field
  couples to the electron density. As shown in Ref.\cite{Gao2020},
  this can be achieved by driving the system with a transverse laser
beam of frequency $\omega_L$ which is detuned from the cavity
frequency by $\delta_c=\omega_c - \omega_L$.  For sufficiently strong
laser driving, the dominant light-matter interaction is induced via
two-photon diamagnetic processes \cite{Gao2020} and reads
\begin{equation}
\label{H_lighmatter}
H_{\rm light-matter} =\sum \limits_{ \vec{k},\sigma} \sum \limits_{\vec{q}=\pm q_0 \hat{x}} g_0~ c^{\dagger}_{\vec{k}+\vec{q},\sigma}c_{\vec{k},\sigma} (b+b^{\dagger}),
\end{equation}
where $b^{\dagger}$ and $c^{\dagger}_{\vec{k},\sigma}$ are the
creation operator of cavity photons of momentum
$\vec{q}_0$ and electrons of momentum $\vec{k}$ and spin $\sigma$
respectively. The coupling strength $g_0$ is tuneable by the intensity
of the external laser beam.
 \begin{figure}[t]
   \centering
  \includegraphics[width=0.48\textwidth]{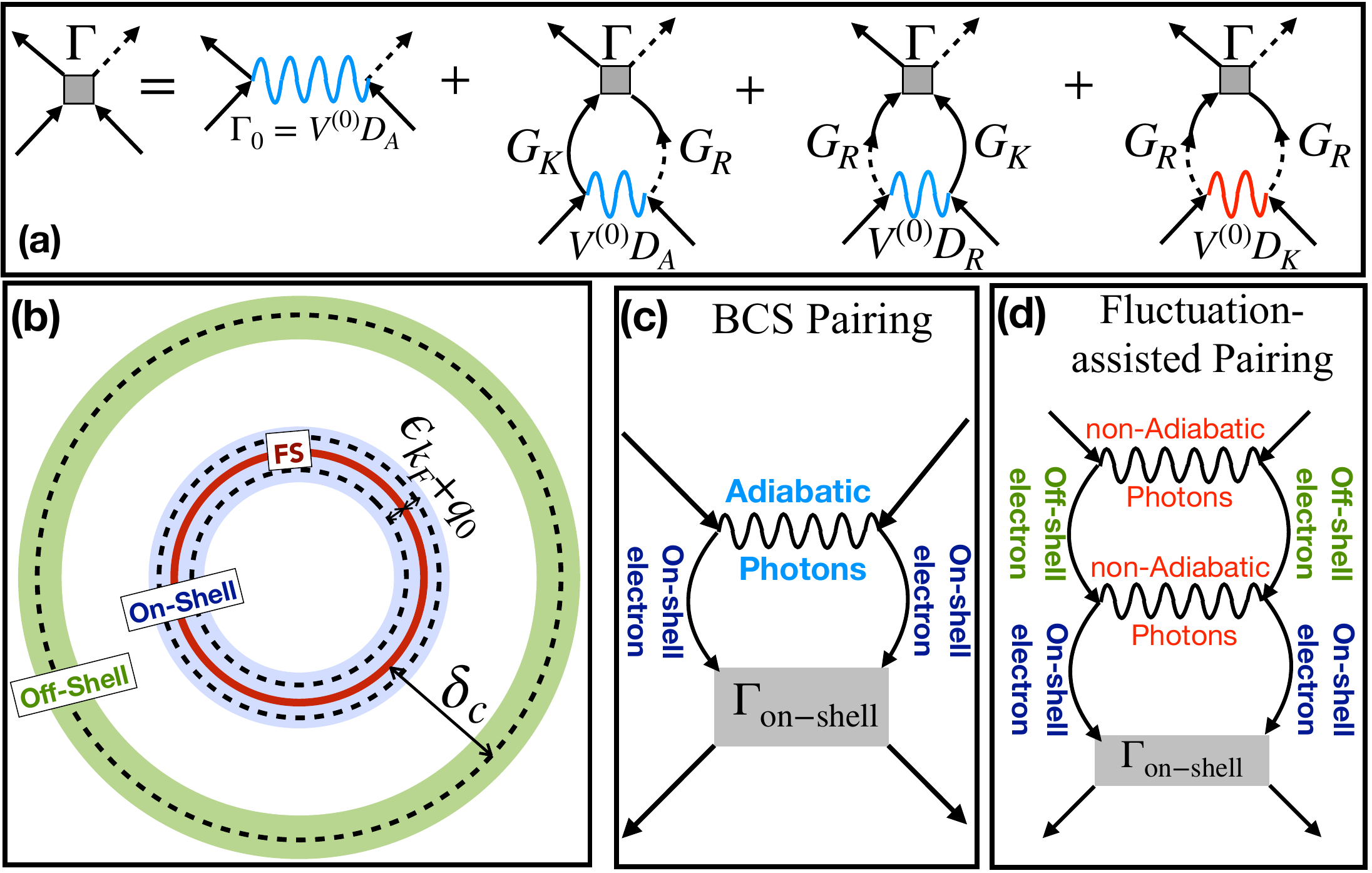}
  \caption{(a) The real-time Keldysh formulation of the Bethe-Salpeter
    equation for the vertex function $\Gamma$ in the pairing
    channel. It contains the bare vertex $\Gamma_0$ (1st diagram), the
    standard BCS terms $\Gamma_{\rm{BCS}}$ (2nd and 3rd diagrams),
    and the ``fluctuation term'', $\Gamma_{\rm{fluct}}$ (4th
    diagram). The latter leads to a non-BCS-type of pairing. (b)
    The electronic energy structure
    can be divided between on-shell electrons in the vicinity of the FS: $\omega_1 \sim \epsilon_{k_F\pm q_0}$ and
    off-shell electrons excited to photonic frequencies: $\omega_1
    \sim \delta_c$ (see
    eq.~\eqref{eq:convolutions}). (c) Standard BCS pairing:
    the scattering via off-shell adiabatic photons leaves the
    electrons on-shell. (d) Fluctuation-induced pairing: on-shell
    non-adiabatic photons at frequencies $\pm \delta_c$ induce pairing between
    on-shell electrons through an intermediate transition to off-shell
    electronic states.} 
   \label{fig:pairing}
   \end{figure}
{\it Causal structure of the pairing.}
We compute the superconducting critical temperature $T_c$ via the
pairing instability of the vertex function $\Gamma$ in the Cooper
pairing channel, involving electrons moving with opposite momenta
$\vec{p}$ and $-\vec{p}$. The Dyson equation for the vertex
function is known as the Bethe-Salpeter
(BS) equation \cite{abrikosov2012methods}, which we formulate using
real-time Green's functions (GFs) defined on the Keldysh closed
time-contour. Besides being suited to include cavity loss (and
incoherent pumping), this approach also allows to clearly separate
the non-BCS-type, fluctuation-induced pairing we propose here from the standard BCS
pairing.
The BS equation in its simplest form has the following structure
(see Fig. \ref{fig:pairing}(a) and Appendix B~\cite{note}),
\begin{equation}
  \Gamma=\Gamma_0+\Gamma_{\rm{BCS}}+\Gamma_{\rm fluct}
  \label{eq:BS}
\end{equation}
where $\Gamma_0=V^{(0)}D_A$ is the bare vertex. Here $\Gamma(\vec{p},\omega)$ is
a function of the relative momentum and frequency of the incoming
electrons. 
Within the ladder approximation \cite{abrikosov2012methods,ladder} for the vertex function we have,
%
  \begin{eqnarray}
  \label{eq:convolutions}
&&\Gamma_{\rm{BCS}}^{A(R)}(p)=\! \! \mathbf{i} \! \!  \int_k  \! \! V^{(0)} {D}_{A(R)}(k-p) ~G_{K(R)}(k) ~G_{R(K)}(-k) \Gamma(k) \nonumber \\
&&\Gamma_{\rm{fluct}}(p) = \mathbf{i} \int_k  V^{(0)} {D}_K(k-p) ~G_R(k) ~G_R(-k) \Gamma(k),
\end{eqnarray}
%
with the coupling function $V^{(0)}(\vec{k}-\vec{p})=g_0^2\delta_c~
\delta_{\vec{k}-\vec{p},\pm q_0\hat{x}}$. Here, $p=(\vec{p},\omega),k=(\vec{k},\omega_1)$ and $ \int_k=  \int d \vec{k} d\omega_1 /(2\pi)^3$. We denote here the electron and
photon GFs by $G$ and $D$, respectively. Within the real-frequency Keldysh
formulation, each GF can be of two types: a retarded (advanced) GF (denoted by the subscript $R (A)$) or a Keldysh
GF (denoted by the subscript $K$).

Formulated in momentum-frequency space, the retarded GF contains only information
about the dispersion and lifetime of the (quasi)particles, while the
Keldysh component explicitly depends also on the occupation of the
quasiparticle modes.
      In thermal equilibrium, these two Green's functions are connected
      by a Fluctuation-Dissipation relation \cite{kamenevbook}. We
      first consider a situation where the whole system is at
      temperature $T$, and the GFs entering the BS equation read \cite{kamenevbook}
      (see appendix A~\cite{note}),
\begin{eqnarray}
\label{Greenbare} 
 G_{R(A)} (\vec{k}, \omega)&=&\frac{1}{\omega-\epsilon_{k} \pm \mathbf{i} 0^+},\nonumber \\
G_{K} (\vec{k},\omega)&=& -2\pi \mathbf{i} \tanh\left(\frac{\omega}{2T}  \right) \delta(\omega-\epsilon_{k}), \nonumber \\
D_{R(A)}(\omega)&=& \frac{1}{2} \frac{1}{(\omega \pm \mathbf{i}0^+)^2-\delta_c^2}, \\
D_K( \omega) &=& \frac{-\pi \mathbf{i}}{2\delta_c} \left[   \delta(\omega - \delta_c) - \delta(\omega + \delta_c) \right] \coth \left( \frac{\omega}{2T} \right). \nonumber
 \end{eqnarray}
 %
 %
Here, $\epsilon_{k}$ is the dispersion of the electrons measured
from the Fermi energy, $E_F$.
Let us now discuss the physical interpretation of the terms
contributing to the BS equation \eqref{eq:BS}.
The standard BCS term $\Gamma_{\rm BCS}= \Gamma_{\rm BCS}^A+
\Gamma_{\rm BCS}^R$ (2nd and 3rd diagram in
Fig.~\ref{fig:pairing}(a)) contains only the retarded (advanced) photon
GF $D_{R(A)}$, while the fluctuation term $\Gamma_{\rm fluct}$ (4th diagram in
Fig.~\ref{fig:pairing}(a)) contains only $D_K$.
Hence, the BCS term $\Gamma_{\rm{BCS}}$ is directly affected only by the dispersion and
lifetime of the photons, but not by the their distribution. On the other
hand, the fluctuation term $\Gamma_{\rm fluct}$ knows about how photon modes are
occupied. Correspondingly, while the BCS vertex contains the electron
$G^K$ proportional to the Fermi-Dirac distribution, the fluctuation
term only contains retarded electron GFs i.e. depends only on the
electron dispersion and lifetime but not directly on their
distribution.


Since the number
of photons is not conserved and in the low-Kelvin regime there is
essentially no thermal occupation of an optical cavity on average, the photons
can be present only due to vacuum fluctuations (or if we include
cavity loss by the corresponding noise, as we shall see later), which explains the
nomenclature $\Gamma_{\rm fluct}$.

{\it Energy structure of the pairing.}
Besides their complementarity in terms of the causal structure
illustrated above, the fluctuation-induced pairing and the BCS pairing
differ sharply in their energy structure.
This can be understood by using the separation of energy scales
between the on-shell and off-shell electrons, shown in
Fig. \ref{fig:pairing}(b). On-shell electrons have frequencies $\omega_1 \sim \epsilon_{k_F \pm q_0}$ (see eq.~\eqref{eq:convolutions}), where $\epsilon_{k_F \pm q_0}$ is a small characteristic
scale i.e
$\epsilon_{k_F \pm q_0} \ll E_F$ as well as  $\epsilon_{k_F \pm q_0} \ll \delta_c$. These low energy
electrons, highlighted by the blue shell in
Fig. \ref{fig:pairing}(b), are the ones that eventually form the Cooper
pairs, as signaled by a divergent pairing amplitude
$\Gamma_{\rm{on-shell}}$ as a solution to the BS equation \eqref{eq:BS} for $T<T_c$. On the
other hand, off-shell electrons (highlighted by the green shell in
Fig. \ref{fig:pairing}(b)) have frequencies close to the photon resonance
frequency i.e. far away from the FS: $\omega_1 \sim
\delta_c$. Off-shell electrons are not the ones actually building the
pair, but can play a crucial role in intermediate scattering
processes, as we shall see in the case when the fluctuation term is included.

{\it BCS pairing.}
As illustrated in Fig. \ref{fig:pairing}(c), here
an adiabatic off-shell photon with
frequency $\omega_1-\omega\sim \epsilon_{k_F\pm q_0} \ll \delta_c$,
scatters an on-shell electron (corresponding to the peak in $G^K$ around $\omega_1\sim \epsilon_{k_F\pm q_0}$)
to a state which is still in the vicinity of the FS. This state is on-shell since the transferred momentum $q_0\ll k_F$. In
eq.~\eqref{eq:convolutions}, $V^{(0)} D_{R(A)} $ in $\Gamma_{\rm BCS}$
can thus be substituted by a negative constant $\sim -g_0^2/2\delta_c $ quantifying the net
attractive interaction, as in the standard BCS scenario involving
phonons. This gives the following equation for $T_c^{\rm{BCS}}$ (see Appendix C for
a detailed derivation~\cite{note})
\begin{equation}
4 \tilde{g} \delta_c  \frac{\tanh \left( \frac{\epsilon_{k_F+q_0}}{2T_c^{\rm{BCS}}} \right) }{\epsilon_{k_F+q_0}} =1.
\label{TcBCS}
\end{equation}
Here, $\tilde{g} =  g_0^2/(4\pi \delta_c)^2$ is the dimensionless coupling. 
We observe that, differently from the standard phonon case, there are no
integrals over loop momentum left due to the long-range nature of the
photon-mediated interaction: $V^{(0)}(\vec{q}) \propto
\delta_{\vec{q},\pm q_0 \hat{x}}$. For $T\gg\epsilon_{k_F+q_0}$, the linear vanishing
$\tanh(x)\simeq x$ of the electron occupation is crucial
in cutting off the $1/\epsilon$ divergence at the FS ($\epsilon_{k_F
  +q_0}\propto q_0$ is finite for finite $q_0$ but is the smallest
scale), yielding $
T_c^{\mathrm{BCS}} \sim  2 \tilde{g} \delta_c
$. 

\begin{figure}[t]
   \centering
  \includegraphics[width=0.48\textwidth]{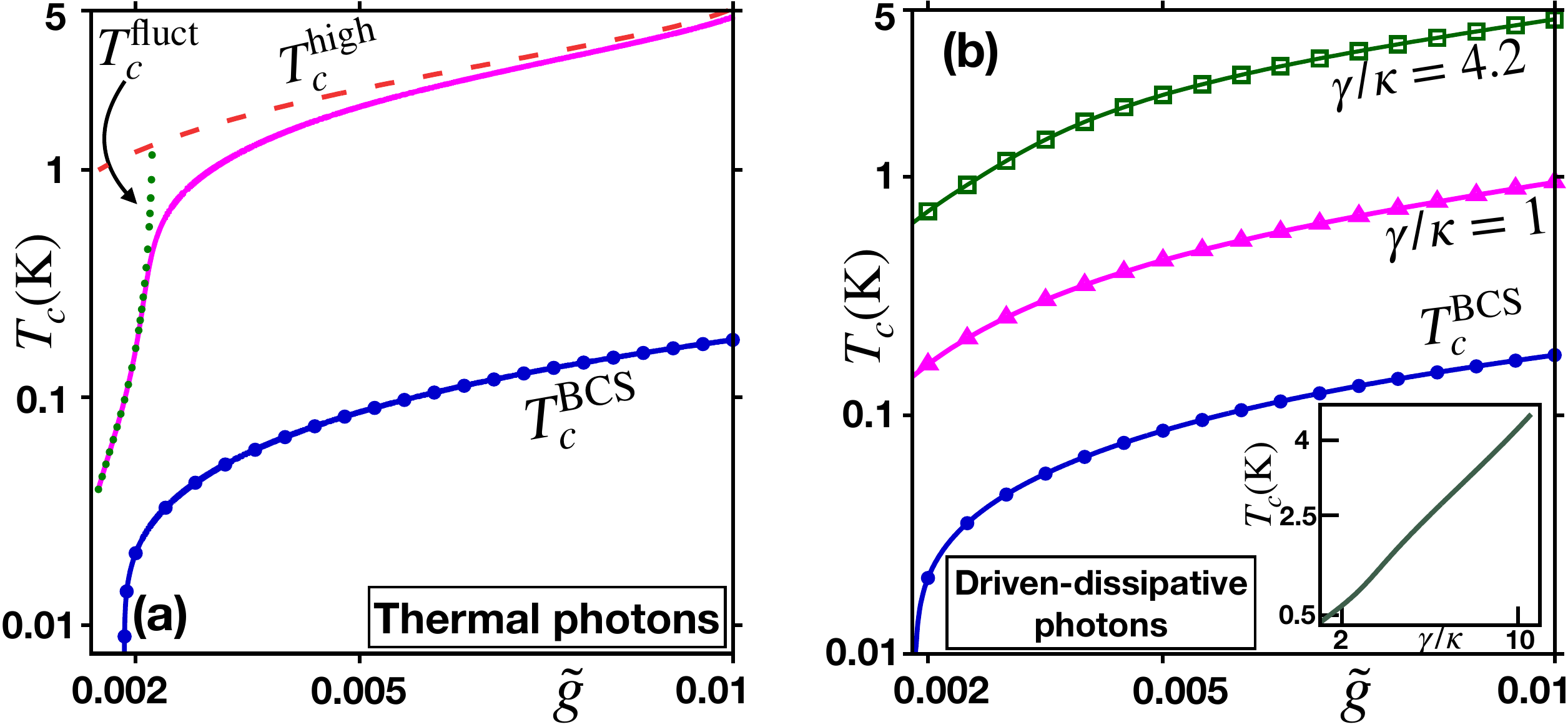}
  \caption{Critical temperature $T_c$ vs dimensionless coupling
    strength $\tilde{g}$ in log plot. (a) For thermal photons, the blue solid line with
    circles corresponds to the prediction \ref{TcBCS}, based on solely the BCS
    pairing (see Fig. \ref{fig:pairing}(c)). The magenta solid
    line corresponds to the prediction which also includes the
    fluctuation-induced pairing (see Fig. \ref{fig:pairing}(d)), as
    well as a finite Fermi-liquid-type electron lifetime (see
    eq.~\eqref{eq:FL_lifetime}).
    The green dotted line shows the prediction for long-lived
    electrons from eq. \ref{TcFluct}, while the red-dashed line
    corresponds to the prediction of eq. \ref{TcHigh} for short-lived
    electrons. (b) For driven dissipative photons, solid line with circles shows the BCS prediction which is almost unaffected by photon loss and incoherent pump. Solid lines with triangles show the fluctuation-assisted enhancement in the purely lossy case $\gamma=\kappa$ which is further amplified by incoherent pumping $\gamma>\kappa$ shown by solid line with squares. Inset (linear plot) shows $T_c$ increases almost linearly with $\gamma$. We choose $\delta_c=0.19\rm{ THz}, \epsilon_{k_F+q_0}/\delta_c=0.007,\kappa/\delta_c=0.01$ and for the inset $\tilde{g}=0.004$.  } 
   \label{fig:TcNumerics}
   \end{figure}

{\it Fluctuation-enhanced pairing.}
On the other hand, photon fluctuations are
concentrated around the cavity resonance frequency, which in the frame
of the driving laser corresponds to $\delta_c$. Hence, in the fluctuation term, pairing is mediated by
on-shell non-adiabatic photons, with $\omega_1-\omega \sim \delta_c$
which scatter electrons off shell. The frequency dependence of the
photon GF $D^K(\omega)\sim \delta(\omega-\delta_c)$ can never be
neglected in eq.~\eqref{eq:convolutions}. Due to the presence of $\Gamma_{\rm
fluct}$, the BS equation \eqref{eq:BS} is not diagonal in frequency,
as it couples the electronic
on-shell and off-shell
sectors. As shown in Fig.\ref{fig:pairing}(d), a further
scattering process where a second non-adiabatic photon brings the electron
back on-shell closes the BS equation. The latter leads to the
following equation for the critical temperature (see Appendix D for
a detailed derivation~\cite{note})
\begin{equation}
  \frac{4 \tilde{g}\delta_c}{\epsilon_{k_F+q_0}}\tanh\left(\frac{\epsilon_{k_F+q_0}}{2T_c^{\rm{fluct}}}\right)+\frac{32}{3} \frac{\tilde{g}^2\delta_c^2}{\epsilon_{k_F+q_0}^2}\left( \coth \frac{\delta_c}{2T_c^{\rm{fluct}}} \right)^2=1.
  \label{TcFluct}
\end{equation}
Note that the BCS
contribution (first term on the L.H.S of eq.~\eqref{TcFluct}) depends
on the electron occupation, while the non-BCS-type contribution
(second term on the L.H.S of eq.~\eqref{TcFluct}) depends on the
photon occupation. 
Fig.~\ref{fig:TcNumerics} (a) shows
that the fluctuation-induced pairing significantly enhances
superconductivity, leading to a critical temperature
$T_c^{\rm{fluct}}$ (dotted green line) increased by an order of magnitude
with respect to the BCS prediction $T_c^{\rm{BCS}}$ (solid line with
circles).
This enhancement can be understood in the following way:  
for not too small laser detunings, the average thermal occupation of cavity
photons is negligible: $\coth(\delta_c/2T)\sim 1$, so that only vacuum
fluctuations remain (later we will discuss the impact of cavity loss
and incoherent pump). This allows to obtain the simple expression for
the critical temperature $T_c^{\mathrm{fluct}} \sim  2 \tilde{g}
\delta_c /[1-32(\tilde{g}\delta_c)^2 /(3\epsilon_{k_F+q_0}^2)]$. For
realistic parameters of 2D materials (LAO/STO) coupled to a split-ring cavity \cite{Gao2020,STO,STOEf}, the ratio
$\delta_c/\epsilon_{k_F+q_0} \sim 135$, and hence the
fluctuation-induced term significantly reduces the denominator from
$1$ already at moderate coupling strengths $\tilde{g}\sim 0.002$,
leading to a significant increase in $T_c$. We also note that the quantities appearing in the
2nd term of eq. (\ref{TcFluct}) are squared due to the additional intermediate scattering to off-shell states
described above. The perturbative expansion in terms of such
scattering processes is controlled as long at
$\tilde{g},\epsilon_{k_F+q_0}/\delta_c\ll 1$ (see Appendix D~\cite{note}).
Moreover, for the high-energy and
  low-momentum photons involved in the non-BCS pairing, in
Ref. \cite{Note1} we show at the non-perturbative level that: a) particle-hole
  excitations only weakly affect the cavity-photon dynamics as well as
  the electron pairing; b) higher-order corrections to the
  electron-photon vertex are small. All these effects
  are thus safely neglected in our BS equation.

At the current level of description, the finite temperature cannot remove the
$1/\epsilon^2$ divergence at the FS arising from the non-BCS-type pairing. 
In order to remove this divergence, we need to take into account the
finite lifetime of electrons. We
consider here a Fermi-liquid scaling of the quasiparticle lifetime in
two dimensions \citep{giuliani1982lifetime,SSarma,RPASelf}, 
\begin{equation}
  \label{eq:FL_lifetime}
\frac{1}{\tau_{e,\rm{cou}}(T;\epsilon_k)}= \frac{\pi}{8}   \frac{{\rm max}
  (T,\epsilon_k)^2}{E_F} \log \left[
  \frac{E_F}{{\rm max} (T,\epsilon_k)} \right],
\end{equation}
induced by the screened Coulomb interaction between on-shell electrons. This
Fermi-liquid lifetime introduces also the leading temperature
dependence of the fluctuation-induced pairing.
This yields the critical temperature indicated by the solid (magenta) line in
Fig. ~\ref{fig:TcNumerics}(a). For a finite cavity wave
vector $q_0$ and at sufficiently low $T$, the quasiparticle energy
$\epsilon_{k_F+q_0}$ dominates over the broadening
$\tau_{e,\rm{cou}}^{-1}(T;\epsilon_k)$, so that $T_c$ closely follows the
prediction $T_c^{\rm{fluct}}$ for infinitely-long-lived
electrons. $T_c$ is raised further by increasing
the coupling $\tilde{g}$, until the temperature becomes large enough
for quasiparticle broadening to become appreciable, leading to the
flattening of the critical-temperature curve as a function of $\tilde{g}$. A large
enhancement induced by photon fluctuations still remains compared to BCS prediction. 
In this regime, $\tau_{e,\rm{cou}}^{-1}(T;\epsilon_k)\gg \epsilon_{k_F+q_0}$ and the critical temperature is approximately determined
by the following equation (see Appendix E for
a detailed derivation~\cite{note}):
%
\begin{equation}
 \frac{2 \tilde{g}\delta_c}{T_c^{\rm{high}}}+ 
 \frac{32\tilde{g}^2\delta_c^2}{\tau_{e,\rm{cou}}^{-2}(T_c^{\mathrm{high}})}\left( \coth \frac{\delta_c}{2T_c^{\rm{high}}} \right)^2  =1.
 \label{TcHigh}
\end{equation}
In the high $T$ limit, $T_c^{\mathrm{high}}$ (dashed red line) shows
good agreement with the full numerical answer. At this point, it is worthwhile to mention
that $T_c^{\mathrm{high}}$ is also valid in the limit of vanishing
cavity wave vector $q_0 \to 0$. Referring to the discussion made in
the model section, we see that the
fluctuation-assisted enhancement of superconductivity thus applies to both
the split-ring-cavity geometry considered in Ref.\cite{Gao2020} and
to the planar microcavity geometry considered in Ref.\cite{Schlawin2019}.

Let us show more explicitly that the non-BCS-type pairing induced by
fluctuations is appreciable only if the interactions are long ranged
i.e. when the bosonic mediator can transfer momenta which are
concentrated in a
narrow window. In order to obtain a simple estimate, we substitute the
delta-function in $V^{(0)}(\vec{q})$ with a box of fixed width. In this case, the factor $1/\tau_{e,\rm{cou}}^{-2}$ in the fluctuation term of
eq.~\eqref{TcHigh} is replaced by
\begin{equation}
-\frac{1}{2W}\int \limits_{-W}^{W} \frac{d\epsilon }{\left( \epsilon  -\frac{\mathbf{i}}{\tau_{e,\rm{cou}}}\right)\left( \epsilon  -\frac{\mathbf{i}}{\tau_{e,\rm{cou}}}\right)} =\frac{1}{W^2+\tau_{e,\rm{cou}}^{-2}},
\end{equation}
%
%
%
where $W$ is the box-width in units of energy.
If the width $W$ exceeds the inverse electron lifetime
$\tau_{e,\rm{cou}}^{-1}$ at the Fermi surface, the fluctuation-induced
enhancement of paring  ($\propto \delta_c^2/W^2$) will be cut off. This explains why the effect discussed in this work
is not relevant for the standard phonon-mediated
pairing. There the energy window $W$ is set by the Debye frequency, which
is large compared to electronic scales. In this regime,
the BCS term takes the known logarithmic form $T_c^{\rm{BCS}} \propto W \exp(-W/(\tilde{g}\delta_c))$.

{\it Impact of photon loss and incoherent pump.}
Photon loss out of the cavity mirrors is unavoidable and typically
happens at an appreciable rate $\gamma_{\rm loss}$. The resulting damping of
photons is introduced in the retarded/advanced GFs $D_{R/A}$ in
Eq. \ref{Greenbare} by substituting $0^+\to\kappa$, while the corresponding noise is included through
the Keldysh GF \cite{Emanuele2013,Sieberer_2016,lang2018nonequilibrium} as
%
%
%
%
\begin{equation}
D_K(\vec{q}, \omega) =- \mathbf{i}\frac{\gamma}{ \delta_c} \frac{\omega^2+\kappa^2+\delta_c^2}{(\omega^2-\kappa^2-\delta_c^2)^2+4\kappa^2\omega^2}.
\end{equation}
Here the parameter $\gamma$ quantifies the noise level. If, apart from
the coherent laser drive, the cavity is not further illuminated, then
$\gamma=\kappa=\gamma_{\rm loss}/2$ i.e. the loss rate sets both the damping and the
noise. We will also study the effect of an incoherent pump (as
resulting from a broadband illumination) at rate $\gamma_{\rm
  pump}<\gamma_{\rm loss}$. In this case the net loss rate becomes $\kappa=(\gamma_{\rm
  loss}-\gamma_{\rm pump})/2$ while the total noise level $\gamma=(\gamma_{\rm
  loss}+\gamma_{\rm pump})/2>\kappa$. For any finite cavity-loss rate,
inelastic electron-photon scattering further reduces the electron
lifetime: $\tau_e^{-1}=\tau_{e, \rm cou}^{-1}+\tau_{e,\rm cav}^{-1}$,
with $\tau_{e,\rm cav}^{-1}\simeq 2\tilde{g}\gamma/(1+\kappa^2/\delta_c^2) $ (see Appendix F~\cite{note}).
Assuming $\delta_c\gg\kappa$ and $\tau_e^{-1}\gg\epsilon_{k_{\rm F}+q_0}$, the equation for the critical temperature takes the
simple form (see Appendix F~\cite{note} for a detailed derivation):
\begin{equation}
 \frac{2 \tilde{g}\delta_c}{T_c^{\rm noise}}+ 
 4\tilde{g}^2\frac{\gamma^2}{\kappa^2}\frac{\delta_c^2}{\tau_e^{-2}(T_c^{\rm noise})}\left[1-\frac{1}{1+\frac{\tau_e^{-1}(T_c^{\rm noise})}{2\kappa}}\right] =1.
 \label{TcNoise}
\end{equation}
This time we used $T_c^{\rm noise}$ as opposed to $T_c^{\rm fluct}$,
since the presence of a on-shell photon is not due to vacuum
fluctuations but to loss-induced noise or, at finite pump rates,  to a
finite average occupation of the cavity mode. We see indeed that the thermal
$\coth$ is replaced here by $\gamma/\kappa=1+2n_{\rm ph}$, with
$n_{\rm ph}$ being the average incoherent occupation of the
cavity. When $\gamma_{\rm
  pump}$ approaches $\gamma_{\rm loss}$ our model needs to be extended
to include pump saturation that prevents the divergence in the photon number. $T_c^{\rm noise}$ is
shown in Fig. \ref{fig:TcNumerics}(b). While the BCS prediction $T_c^{\rm{BCS}}\simeq 2\tilde{g}\delta_c/(1+\kappa^2/\delta_c^2)  $ (Fig. \ref{fig:TcNumerics}(b) with circles)
remains essentially unaffected by loss (for $\delta_c\gg\kappa$) and
does not depend on incoherent pumping, the non-BCS-type pairing still
provides a strong enhancement of superconductivity  (Fig. \ref{fig:TcNumerics}(b) with triangles), which is further amplified by a
finite incoherent-pump rate  (Fig. \ref{fig:TcNumerics}(b) with squares). By comparing eq.~\eqref{TcNoise} with the
closed-system expression, we see that for
$\kappa\neq 0$ the critical temperature is reduced by the second term
in the square bracket. Still, for $\kappa\tau_e\ll 1$ the reduction is
negligible. Moreover, by increasing the
incoherent-pump rate $\gamma$ we can further increase $T_c$ almost
linearly, as shown in the inset of
Fig. \ref{fig:TcNumerics}(b). Hence, with increasing $\gamma$,
$\tau_{e,\rm cou}^{-1}$ increases faster ($\propto T^2 \log T$) than
the linearly increasing $\tau_{e,\rm cav}^{-1}$. The Coulomb
lifetime thus still serves as the dominant quasiparticle-damping
process.For this reason, the photon-induced redistribution of
  quasiparticles, predicted to be appreciable in certain cavity setups
  \cite{Curtis2019} and observed by proper irradiation \cite{klapwijk2020discovery,tikhonov2020microwave}, can instead be neglected in our case. We can predict an increase in $T_c$ as long as the
Fermi-liquid behavior \eqref{eq:FL_lifetime} still holds, which sets
$T_c$ in the range of the Fermi temperature $T_F$ ($\sim 13 $K
for the materials considered in Ref. \cite{Gao2020}).



{\it Conclusions.}
We have shown that cavity-mediated interactions between electrons induce a
non-BSC-type of pairing mechanism triggered by non-adiabatic photon fluctuations,
which can largely enhance the superconducting critical temperature $T_c$
under realistic conditions. The characteristic non-BCS dependence of $T_c$ on the tuneable light-matter coupling could be experimentally observed. Moreover, an increase of Tc with the number incoherent photons in the cavity would be indicative of the present non-BCS pairing to be active. These features could be also theoretically reproduced using quantum Monte Carlo for fermion-boson systems\cite{QMC}, proper extensions to long-range interactions of matrix product states\cite{MPS} or density functional theory \cite{DFT}, as well as finite-frequency extensions of functional renormalization group technique \cite{Karrasch_2008}.


%
\begin{acknowledgments}
We thank Bernhard Frank, Dieter Jaksch, Johannes Lang, Andrew Millis,
Frank Schlawin, and Michael Sentef for useful discussions. 
\end{acknowledgments}

\bibliographystyle{apsrev4-1}
\bibliography{LIS.bib}

\begin{thebibliography}{44}%
\makeatletter
\providecommand \@ifxundefined [1]{%
 \@ifx{#1\undefined}
}%
\providecommand \@ifnum [1]{%
 \ifnum #1\expandafter \@firstoftwo
 \else \expandafter \@secondoftwo
 \fi
}%
\providecommand \@ifx [1]{%
 \ifx #1\expandafter \@firstoftwo
 \else \expandafter \@secondoftwo
 \fi
}%
\providecommand \natexlab [1]{#1}%
\providecommand \enquote  [1]{``#1''}%
\providecommand \bibnamefont  [1]{#1}%
\providecommand \bibfnamefont [1]{#1}%
\providecommand \citenamefont [1]{#1}%
\providecommand \href@noop [0]{\@secondoftwo}%
\providecommand \href [0]{\begingroup \@sanitize@url \@href}%
\providecommand \@href[1]{\@@startlink{#1}\@@href}%
\providecommand \@@href[1]{\endgroup#1\@@endlink}%
\providecommand \@sanitize@url [0]{\catcode `\\12\catcode `\$12\catcode
  `\&12\catcode `\#12\catcode `\^12\catcode `\_12\catcode `\%12\relax}%
\providecommand \@@startlink[1]{}%
\providecommand \@@endlink[0]{}%
\providecommand \url  [0]{\begingroup\@sanitize@url \@url }%
\providecommand \@url [1]{\endgroup\@href {#1}{\urlprefix }}%
\providecommand \urlprefix  [0]{URL }%
\providecommand \Eprint [0]{\href }%
\providecommand \doibase [0]{http://dx.doi.org/}%
\providecommand \selectlanguage [0]{\@gobble}%
\providecommand \bibinfo  [0]{\@secondoftwo}%
\providecommand \bibfield  [0]{\@secondoftwo}%
\providecommand \translation [1]{[#1]}%
\providecommand \BibitemOpen [0]{}%
\providecommand \bibitemStop [0]{}%
\providecommand \bibitemNoStop [0]{.\EOS\space}%
\providecommand \EOS [0]{\spacefactor3000\relax}%
\providecommand \BibitemShut  [1]{\csname bibitem#1\endcsname}%
\let\auto@bib@innerbib\@empty
\bibitem [{\citenamefont {Paravicini-Bagliani}\ \emph
  {et~al.}(2019)\citenamefont {Paravicini-Bagliani}, \citenamefont
  {Appugliese}, \citenamefont {Richter}, \citenamefont {Valmorra},
  \citenamefont {Keller}, \citenamefont {Beck}, \citenamefont {Bartolo},
  \citenamefont {R{\"{o}}ssler}, \citenamefont {Ihn}, \citenamefont {Ensslin},
  \citenamefont {Ciuti}, \citenamefont {Scalari},\ and\ \citenamefont
  {Faist}}]{Paravicini-Bagliani2019}%
  \BibitemOpen
  \bibfield  {author} {\bibinfo {author} {\bibfnamefont {G.~L.}\ \bibnamefont
  {Paravicini-Bagliani}}, \bibinfo {author} {\bibfnamefont {F.}~\bibnamefont
  {Appugliese}}, \bibinfo {author} {\bibfnamefont {E.}~\bibnamefont {Richter}},
  \bibinfo {author} {\bibfnamefont {F.}~\bibnamefont {Valmorra}}, \bibinfo
  {author} {\bibfnamefont {J.}~\bibnamefont {Keller}}, \bibinfo {author}
  {\bibfnamefont {M.}~\bibnamefont {Beck}}, \bibinfo {author} {\bibfnamefont
  {N.}~\bibnamefont {Bartolo}}, \bibinfo {author} {\bibfnamefont
  {C.}~\bibnamefont {R{\"{o}}ssler}}, \bibinfo {author} {\bibfnamefont
  {T.}~\bibnamefont {Ihn}}, \bibinfo {author} {\bibfnamefont {K.}~\bibnamefont
  {Ensslin}}, \bibinfo {author} {\bibfnamefont {C.}~\bibnamefont {Ciuti}},
  \bibinfo {author} {\bibfnamefont {G.}~\bibnamefont {Scalari}}, \ and\
  \bibinfo {author} {\bibfnamefont {J.}~\bibnamefont {Faist}},\ }\href
  {\doibase 10.1038/s41567-018-0346-y} {\bibfield  {journal} {\bibinfo
  {journal} {Nature Physics}\ }\textbf {\bibinfo {volume} {15}},\ \bibinfo
  {pages} {186} (\bibinfo {year} {2019})}\BibitemShut {NoStop}%
\bibitem [{\citenamefont {Thomas}\ \emph {et~al.}(2019)\citenamefont {Thomas},
  \citenamefont {Devaux}, \citenamefont {Nagarajan}, \citenamefont {Chervy},
  \citenamefont {Seidel}, \citenamefont {Hagenm{\"{u}}ller}, \citenamefont
  {Sch{\"{u}}tz}, \citenamefont {Schachenmayer}, \citenamefont {Genet},
  \citenamefont {Pupillo},\ and\ \citenamefont {Ebbesen}}]{Thomas2019}%
  \BibitemOpen
  \bibfield  {author} {\bibinfo {author} {\bibfnamefont {A.}~\bibnamefont
  {Thomas}}, \bibinfo {author} {\bibfnamefont {E.}~\bibnamefont {Devaux}},
  \bibinfo {author} {\bibfnamefont {K.}~\bibnamefont {Nagarajan}}, \bibinfo
  {author} {\bibfnamefont {T.}~\bibnamefont {Chervy}}, \bibinfo {author}
  {\bibfnamefont {M.}~\bibnamefont {Seidel}}, \bibinfo {author} {\bibfnamefont
  {D.}~\bibnamefont {Hagenm{\"{u}}ller}}, \bibinfo {author} {\bibfnamefont
  {S.}~\bibnamefont {Sch{\"{u}}tz}}, \bibinfo {author} {\bibfnamefont
  {J.}~\bibnamefont {Schachenmayer}}, \bibinfo {author} {\bibfnamefont
  {C.}~\bibnamefont {Genet}}, \bibinfo {author} {\bibfnamefont
  {G.}~\bibnamefont {Pupillo}}, \ and\ \bibinfo {author} {\bibfnamefont
  {T.~W.}\ \bibnamefont {Ebbesen}},\ }\href {http://arxiv.org/abs/1911.01459}
  {\  (\bibinfo {year} {2019})},\ \Eprint {http://arxiv.org/abs/1911.01459}
  {arXiv:1911.01459} \BibitemShut {NoStop}%
\bibitem [{\citenamefont {Sentef}\ \emph {et~al.}(2018)\citenamefont {Sentef},
  \citenamefont {Ruggenthaler},\ and\ \citenamefont {Rubio}}]{Sentef2018}%
  \BibitemOpen
  \bibfield  {author} {\bibinfo {author} {\bibfnamefont {M.~A.}\ \bibnamefont
  {Sentef}}, \bibinfo {author} {\bibfnamefont {M.}~\bibnamefont
  {Ruggenthaler}}, \ and\ \bibinfo {author} {\bibfnamefont {A.}~\bibnamefont
  {Rubio}},\ }\href {\doibase 10.1126/sciadv.aau6969} {\bibfield  {journal}
  {\bibinfo  {journal} {Science Advances}\ }\textbf {\bibinfo {volume} {4}},\
  \bibinfo {pages} {eaau6969} (\bibinfo {year} {2018})}\BibitemShut {NoStop}%
\bibitem [{\citenamefont {Curtis}\ \emph {et~al.}(2019)\citenamefont {Curtis},
  \citenamefont {Raines}, \citenamefont {Allocca}, \citenamefont {Hafezi},\
  and\ \citenamefont {Galitski}}]{Curtis2019}%
  \BibitemOpen
  \bibfield  {author} {\bibinfo {author} {\bibfnamefont {J.~B.}\ \bibnamefont
  {Curtis}}, \bibinfo {author} {\bibfnamefont {Z.~M.}\ \bibnamefont {Raines}},
  \bibinfo {author} {\bibfnamefont {A.~A.}\ \bibnamefont {Allocca}}, \bibinfo
  {author} {\bibfnamefont {M.}~\bibnamefont {Hafezi}}, \ and\ \bibinfo {author}
  {\bibfnamefont {V.~M.}\ \bibnamefont {Galitski}},\ }\href {\doibase
  10.1103/PhysRevLett.122.167002} {\bibfield  {journal} {\bibinfo  {journal}
  {Physical Review Letters}\ }\textbf {\bibinfo {volume} {122}},\ \bibinfo
  {pages} {167002} (\bibinfo {year} {2019})}\BibitemShut {NoStop}%
\bibitem [{\citenamefont {Mazza}\ and\ \citenamefont
  {Georges}(2019)}]{Mazza2019}%
  \BibitemOpen
  \bibfield  {author} {\bibinfo {author} {\bibfnamefont {G.}~\bibnamefont
  {Mazza}}\ and\ \bibinfo {author} {\bibfnamefont {A.}~\bibnamefont
  {Georges}},\ }\href {\doibase 10.1103/PhysRevLett.122.017401} {\bibfield
  {journal} {\bibinfo  {journal} {Physical Review Letters}\ }\textbf {\bibinfo
  {volume} {122}},\ \bibinfo {pages} {17401} (\bibinfo {year}
  {2019})}\BibitemShut {NoStop}%
\bibitem [{\citenamefont {Kiffner}\ \emph {et~al.}(2019)\citenamefont
  {Kiffner}, \citenamefont {Coulthard}, \citenamefont {Schlawin}, \citenamefont
  {Ardavan},\ and\ \citenamefont {Jaksch}}]{Kiffner2019}%
  \BibitemOpen
  \bibfield  {author} {\bibinfo {author} {\bibfnamefont {M.}~\bibnamefont
  {Kiffner}}, \bibinfo {author} {\bibfnamefont {J.~R.}\ \bibnamefont
  {Coulthard}}, \bibinfo {author} {\bibfnamefont {F.}~\bibnamefont {Schlawin}},
  \bibinfo {author} {\bibfnamefont {A.}~\bibnamefont {Ardavan}}, \ and\
  \bibinfo {author} {\bibfnamefont {D.}~\bibnamefont {Jaksch}},\ }\href
  {\doibase 10.1103/PhysRevB.99.085116} {\bibfield  {journal} {\bibinfo
  {journal} {Physical Review B}\ }\textbf {\bibinfo {volume} {99}},\ \bibinfo
  {pages} {085116} (\bibinfo {year} {2019})}\BibitemShut {NoStop}%
\bibitem [{\citenamefont {Schlawin}\ \emph {et~al.}(2019)\citenamefont
  {Schlawin}, \citenamefont {Cavalleri},\ and\ \citenamefont
  {Jaksch}}]{Schlawin2019}%
  \BibitemOpen
  \bibfield  {author} {\bibinfo {author} {\bibfnamefont {F.}~\bibnamefont
  {Schlawin}}, \bibinfo {author} {\bibfnamefont {A.}~\bibnamefont {Cavalleri}},
  \ and\ \bibinfo {author} {\bibfnamefont {D.}~\bibnamefont {Jaksch}},\ }\href
  {\doibase 10.1103/PhysRevLett.122.133602} {\bibfield  {journal} {\bibinfo
  {journal} {Physical Review Letters}\ }\textbf {\bibinfo {volume} {122}},\
  \bibinfo {pages} {133602} (\bibinfo {year} {2019})}\BibitemShut {NoStop}%
\bibitem [{\citenamefont {Allocca}\ \emph {et~al.}(2019)\citenamefont
  {Allocca}, \citenamefont {Raines}, \citenamefont {Curtis},\ and\
  \citenamefont {Galitski}}]{Allocca2019}%
  \BibitemOpen
  \bibfield  {author} {\bibinfo {author} {\bibfnamefont {A.~A.}\ \bibnamefont
  {Allocca}}, \bibinfo {author} {\bibfnamefont {Z.~M.}\ \bibnamefont {Raines}},
  \bibinfo {author} {\bibfnamefont {J.~B.}\ \bibnamefont {Curtis}}, \ and\
  \bibinfo {author} {\bibfnamefont {V.~M.}\ \bibnamefont {Galitski}},\ }\href
  {\doibase 10.1103/PhysRevB.99.020504} {\bibfield  {journal} {\bibinfo
  {journal} {Phys. Rev. B}\ }\textbf {\bibinfo {volume} {99}},\ \bibinfo
  {pages} {020504} (\bibinfo {year} {2019})}\BibitemShut {NoStop}%
\bibitem [{\citenamefont {Gao}\ \emph {et~al.}(2020)\citenamefont {Gao},
  \citenamefont {Schlawin}, \citenamefont {Buzzi}, \citenamefont {Cavalleri},\
  and\ \citenamefont {Jaksch}}]{Gao2020}%
  \BibitemOpen
  \bibfield  {author} {\bibinfo {author} {\bibfnamefont {H.}~\bibnamefont
  {Gao}}, \bibinfo {author} {\bibfnamefont {F.}~\bibnamefont {Schlawin}},
  \bibinfo {author} {\bibfnamefont {M.}~\bibnamefont {Buzzi}}, \bibinfo
  {author} {\bibfnamefont {A.}~\bibnamefont {Cavalleri}}, \ and\ \bibinfo
  {author} {\bibfnamefont {D.}~\bibnamefont {Jaksch}},\ }\href {\doibase
  10.1103/PhysRevLett.125.053602} {\bibfield  {journal} {\bibinfo  {journal}
  {Phys. Rev. Lett.}\ }\textbf {\bibinfo {volume} {125}},\ \bibinfo {pages}
  {053602} (\bibinfo {year} {2020})}\BibitemShut {NoStop}%
\bibitem [{\citenamefont {Ashida}\ \emph {et~al.}(2020)\citenamefont {Ashida},
  \citenamefont {Imamoglu}, \citenamefont {Faist}, \citenamefont {Jaksch},
  \citenamefont {Cavalleri},\ and\ \citenamefont {Demler}}]{Ashida2020}%
  \BibitemOpen
  \bibfield  {author} {\bibinfo {author} {\bibfnamefont {Y.}~\bibnamefont
  {Ashida}}, \bibinfo {author} {\bibfnamefont {A.}~\bibnamefont {Imamoglu}},
  \bibinfo {author} {\bibfnamefont {J.}~\bibnamefont {Faist}}, \bibinfo
  {author} {\bibfnamefont {D.}~\bibnamefont {Jaksch}}, \bibinfo {author}
  {\bibfnamefont {A.}~\bibnamefont {Cavalleri}}, \ and\ \bibinfo {author}
  {\bibfnamefont {E.}~\bibnamefont {Demler}},\ }\href
  {http://arxiv.org/abs/2003.13695} {\  (\bibinfo {year} {2020})},\ \Eprint
  {http://arxiv.org/abs/2003.13695} {arXiv:2003.13695} \BibitemShut {NoStop}%
\bibitem [{\citenamefont {Sentef}\ \emph {et~al.}(2020)\citenamefont {Sentef},
  \citenamefont {Li}, \citenamefont {K{\"{u}}nzel},\ and\ \citenamefont
  {Eckstein}}]{Sentef2020}%
  \BibitemOpen
  \bibfield  {author} {\bibinfo {author} {\bibfnamefont {M.~A.}\ \bibnamefont
  {Sentef}}, \bibinfo {author} {\bibfnamefont {J.}~\bibnamefont {Li}}, \bibinfo
  {author} {\bibfnamefont {F.}~\bibnamefont {K{\"{u}}nzel}}, \ and\ \bibinfo
  {author} {\bibfnamefont {M.}~\bibnamefont {Eckstein}},\ }\href {\doibase
  10.1103/PhysRevResearch.2.033033} {\bibfield  {journal} {\bibinfo  {journal}
  {Physical Review Research}\ }\textbf {\bibinfo {volume} {2}},\ \bibinfo
  {pages} {033033} (\bibinfo {year} {2020})}\BibitemShut {NoStop}%
\bibitem [{\citenamefont {Piazza}\ and\ \citenamefont
  {Strack}(2014{\natexlab{a}})}]{Piazza2014Umklapp}%
  \BibitemOpen
  \bibfield  {author} {\bibinfo {author} {\bibfnamefont {F.}~\bibnamefont
  {Piazza}}\ and\ \bibinfo {author} {\bibfnamefont {P.}~\bibnamefont
  {Strack}},\ }\href {\doibase 10.1103/PhysRevLett.112.143003} {\bibfield
  {journal} {\bibinfo  {journal} {Physical Review Letters}\ }\textbf {\bibinfo
  {volume} {112}},\ \bibinfo {pages} {143003} (\bibinfo {year}
  {2014}{\natexlab{a}})}\BibitemShut {NoStop}%
\bibitem [{\citenamefont {Keeling}\ \emph {et~al.}(2014)\citenamefont
  {Keeling}, \citenamefont {Bhaseen},\ and\ \citenamefont
  {Simons}}]{Keeling2014Fermionic}%
  \BibitemOpen
  \bibfield  {author} {\bibinfo {author} {\bibfnamefont {J.}~\bibnamefont
  {Keeling}}, \bibinfo {author} {\bibfnamefont {M.~J.}\ \bibnamefont
  {Bhaseen}}, \ and\ \bibinfo {author} {\bibfnamefont {B.~D.}\ \bibnamefont
  {Simons}},\ }\href {\doibase 10.1103/PhysRevLett.112.143002} {\bibfield
  {journal} {\bibinfo  {journal} {Physical Review Letters}\ }\textbf {\bibinfo
  {volume} {112}},\ \bibinfo {pages} {143002} (\bibinfo {year}
  {2014})}\BibitemShut {NoStop}%
\bibitem [{\citenamefont {Chen}\ \emph {et~al.}(2014)\citenamefont {Chen},
  \citenamefont {Yu},\ and\ \citenamefont {Zhai}}]{Chen2014Superradiance}%
  \BibitemOpen
  \bibfield  {author} {\bibinfo {author} {\bibfnamefont {Y.}~\bibnamefont
  {Chen}}, \bibinfo {author} {\bibfnamefont {Z.}~\bibnamefont {Yu}}, \ and\
  \bibinfo {author} {\bibfnamefont {H.}~\bibnamefont {Zhai}},\ }\href {\doibase
  10.1103/PhysRevLett.112.143004} {\bibfield  {journal} {\bibinfo  {journal}
  {Physical Review Letters}\ }\textbf {\bibinfo {volume} {112}},\ \bibinfo
  {pages} {143004} (\bibinfo {year} {2014})}\BibitemShut {NoStop}%
\bibitem [{\citenamefont {Piazza}\ and\ \citenamefont
  {Strack}(2014{\natexlab{b}})}]{Piazza2014Quantum}%
  \BibitemOpen
  \bibfield  {author} {\bibinfo {author} {\bibfnamefont {F.}~\bibnamefont
  {Piazza}}\ and\ \bibinfo {author} {\bibfnamefont {P.}~\bibnamefont
  {Strack}},\ }\href {\doibase 10.1103/PhysRevA.90.043823} {\bibfield
  {journal} {\bibinfo  {journal} {Physical Review A}\ }\textbf {\bibinfo
  {volume} {90}},\ \bibinfo {pages} {043823} (\bibinfo {year}
  {2014}{\natexlab{b}})}\BibitemShut {NoStop}%
\bibitem [{\citenamefont {Pan}\ \emph {et~al.}(2015)\citenamefont {Pan},
  \citenamefont {Liu}, \citenamefont {Zhang}, \citenamefont {Yi},\ and\
  \citenamefont {Guo}}]{Pan2015Topological}%
  \BibitemOpen
  \bibfield  {author} {\bibinfo {author} {\bibfnamefont {J.-S.}\ \bibnamefont
  {Pan}}, \bibinfo {author} {\bibfnamefont {X.-J.}\ \bibnamefont {Liu}},
  \bibinfo {author} {\bibfnamefont {W.}~\bibnamefont {Zhang}}, \bibinfo
  {author} {\bibfnamefont {W.}~\bibnamefont {Yi}}, \ and\ \bibinfo {author}
  {\bibfnamefont {G.-C.}\ \bibnamefont {Guo}},\ }\href {\doibase
  10.1103/PhysRevLett.115.045303} {\bibfield  {journal} {\bibinfo  {journal}
  {Physical Review Letters}\ }\textbf {\bibinfo {volume} {115}},\ \bibinfo
  {pages} {045303} (\bibinfo {year} {2015})}\BibitemShut {NoStop}%
\bibitem [{\citenamefont {Zheng}\ and\ \citenamefont
  {Cooper}(2016)}]{Zheng2016Superradiance}%
  \BibitemOpen
  \bibfield  {author} {\bibinfo {author} {\bibfnamefont {W.}~\bibnamefont
  {Zheng}}\ and\ \bibinfo {author} {\bibfnamefont {N.~R.}\ \bibnamefont
  {Cooper}},\ }\href {\doibase 10.1103/PhysRevLett.117.175302} {\bibfield
  {journal} {\bibinfo  {journal} {Physical Review Letters}\ }\textbf {\bibinfo
  {volume} {117}},\ \bibinfo {pages} {175302} (\bibinfo {year} {2016})},\
  \Eprint {http://arxiv.org/abs/1604.06630} {arXiv:1604.06630} \BibitemShut
  {NoStop}%
\bibitem [{\citenamefont {Kollath}\ \emph {et~al.}(2016)\citenamefont
  {Kollath}, \citenamefont {Sheikhan}, \citenamefont {Wolff},\ and\
  \citenamefont {Brennecke}}]{Kollath2016Ultracold}%
  \BibitemOpen
  \bibfield  {author} {\bibinfo {author} {\bibfnamefont {C.}~\bibnamefont
  {Kollath}}, \bibinfo {author} {\bibfnamefont {A.}~\bibnamefont {Sheikhan}},
  \bibinfo {author} {\bibfnamefont {S.}~\bibnamefont {Wolff}}, \ and\ \bibinfo
  {author} {\bibfnamefont {F.}~\bibnamefont {Brennecke}},\ }\href {\doibase
  10.1103/PhysRevLett.116.060401} {\bibfield  {journal} {\bibinfo  {journal}
  {Physical Review Letters}\ }\textbf {\bibinfo {volume} {116}},\ \bibinfo
  {pages} {060401} (\bibinfo {year} {2016})},\ \Eprint
  {http://arxiv.org/abs/1502.01817} {arXiv:1502.01817} \BibitemShut {NoStop}%
\bibitem [{\citenamefont {Mivehvar}\ \emph {et~al.}(2017)\citenamefont
  {Mivehvar}, \citenamefont {Ritsch},\ and\ \citenamefont
  {Piazza}}]{Mivehvar2017Superradiant}%
  \BibitemOpen
  \bibfield  {author} {\bibinfo {author} {\bibfnamefont {F.}~\bibnamefont
  {Mivehvar}}, \bibinfo {author} {\bibfnamefont {H.}~\bibnamefont {Ritsch}}, \
  and\ \bibinfo {author} {\bibfnamefont {F.}~\bibnamefont {Piazza}},\ }\href
  {\doibase 10.1103/PhysRevLett.118.073602} {\bibfield  {journal} {\bibinfo
  {journal} {Physical Review Letters}\ }\textbf {\bibinfo {volume} {118}},\
  \bibinfo {pages} {073602} (\bibinfo {year} {2017})}\BibitemShut {NoStop}%
\bibitem [{\citenamefont {Colella}\ \emph {et~al.}(2018)\citenamefont
  {Colella}, \citenamefont {Citro}, \citenamefont {Barsanti}, \citenamefont
  {Rossini},\ and\ \citenamefont {Chiofalo}}]{Colella2018Quantum}%
  \BibitemOpen
  \bibfield  {author} {\bibinfo {author} {\bibfnamefont {E.}~\bibnamefont
  {Colella}}, \bibinfo {author} {\bibfnamefont {R.}~\bibnamefont {Citro}},
  \bibinfo {author} {\bibfnamefont {M.}~\bibnamefont {Barsanti}}, \bibinfo
  {author} {\bibfnamefont {D.}~\bibnamefont {Rossini}}, \ and\ \bibinfo
  {author} {\bibfnamefont {M.-L.}\ \bibnamefont {Chiofalo}},\ }\href {\doibase
  10.1103/PhysRevB.97.134502} {\bibfield  {journal} {\bibinfo  {journal}
  {Physical Review B}\ }\textbf {\bibinfo {volume} {97}},\ \bibinfo {pages}
  {134502} (\bibinfo {year} {2018})}\BibitemShut {NoStop}%
\bibitem [{\citenamefont {Mivehvar}\ \emph {et~al.}(2019)\citenamefont
  {Mivehvar}, \citenamefont {Ritsch},\ and\ \citenamefont
  {Piazza}}]{Mivehvar2019Cavity}%
  \BibitemOpen
  \bibfield  {author} {\bibinfo {author} {\bibfnamefont {F.}~\bibnamefont
  {Mivehvar}}, \bibinfo {author} {\bibfnamefont {H.}~\bibnamefont {Ritsch}}, \
  and\ \bibinfo {author} {\bibfnamefont {F.}~\bibnamefont {Piazza}},\ }\href
  {\doibase 10.1103/PhysRevLett.122.113603} {\bibfield  {journal} {\bibinfo
  {journal} {Physical Review Letters}\ }\textbf {\bibinfo {volume} {122}},\
  \bibinfo {pages} {113603} (\bibinfo {year} {2019})}\BibitemShut {NoStop}%
\bibitem [{\citenamefont {Schlawin}\ and\ \citenamefont
  {Jaksch}(2019)}]{Schlawin2019a}%
  \BibitemOpen
  \bibfield  {author} {\bibinfo {author} {\bibfnamefont {F.}~\bibnamefont
  {Schlawin}}\ and\ \bibinfo {author} {\bibfnamefont {D.}~\bibnamefont
  {Jaksch}},\ }\href {\doibase 10.1103/PhysRevLett.123.133601} {\bibfield
  {journal} {\bibinfo  {journal} {Physical Review Letters}\ }\textbf {\bibinfo
  {volume} {123}},\ \bibinfo {pages} {133601} (\bibinfo {year}
  {2019})}\BibitemShut {NoStop}%
\bibitem [{\citenamefont {Abrikosov}\ \emph {et~al.}(1975)\citenamefont
  {Abrikosov}, \citenamefont {Dzyaloshinskii}, \citenamefont {Gorkov},\ and\
  \citenamefont {Silverman}}]{abrikosov2012methods}%
  \BibitemOpen
  \bibfield  {author} {\bibinfo {author} {\bibfnamefont {A.~A.}\ \bibnamefont
  {Abrikosov}}, \bibinfo {author} {\bibfnamefont {I.}~\bibnamefont
  {Dzyaloshinskii}}, \bibinfo {author} {\bibfnamefont {L.~P.}\ \bibnamefont
  {Gorkov}}, \ and\ \bibinfo {author} {\bibfnamefont {R.~A.}\ \bibnamefont
  {Silverman}},\ }\href@noop {} {\emph {\bibinfo {title} {{Methods of quantum
  field theory in statistical physics}}}}\ (\bibinfo  {publisher} {Dover},\
  \bibinfo {address} {New York, NY},\ \bibinfo {year} {1975})\BibitemShut
  {NoStop}%
\bibitem [{not()}]{note}%
  \BibitemOpen
  \href@noop {} {}\bibinfo {note} {See Supplemental Material for the derivation
  of the effective action using the Schwinger-Keldysh field theoretic
  technique, the solution of the resultant Bethe-Salpeter equation for the
  vertex-function to find critical temperature including both the standard BCS
  and the new non-BCS mechanisms, a discussion on the effect of finite
  life-time of the electrons and the role cavity-losses and incoherent pumping,
  which includes Refs. \cite{Negele_Orland,altland,FermiLiquid}}\BibitemShut
  {NoStop}%
\bibitem [{\citenamefont {Mattuck}(1992)}]{ladder}%
  \BibitemOpen
  \bibfield  {author} {\bibinfo {author} {\bibfnamefont {R.~D.}\ \bibnamefont
  {Mattuck}},\ }\href@noop {} {\emph {\bibinfo {title} {Methods of Quantum
  Field Theory in Statistical Physics}}}\ (\bibinfo  {publisher} {Dover
  Publications, INC. New York},\ \bibinfo {year} {1992})\BibitemShut {NoStop}%
\bibitem [{\citenamefont {Kamenev}(2011)}]{kamenevbook}%
  \BibitemOpen
  \bibfield  {author} {\bibinfo {author} {\bibfnamefont {A.}~\bibnamefont
  {Kamenev}},\ }\href@noop {} {\emph {\bibinfo {title} {Field theory of
  non-equilibrium systems}}}\ (\bibinfo  {publisher} {Cambridge University
  Press},\ \bibinfo {year} {2011})\BibitemShut {NoStop}%
\bibitem [{\citenamefont {Sakudo}\ and\ \citenamefont {Unoki}(1971)}]{STO}%
  \BibitemOpen
  \bibfield  {author} {\bibinfo {author} {\bibfnamefont {T.}~\bibnamefont
  {Sakudo}}\ and\ \bibinfo {author} {\bibfnamefont {H.}~\bibnamefont {Unoki}},\
  }\href {\doibase 10.1103/PhysRevLett.26.851} {\bibfield  {journal} {\bibinfo
  {journal} {Phys. Rev. Lett.}\ }\textbf {\bibinfo {volume} {26}},\ \bibinfo
  {pages} {851} (\bibinfo {year} {1971})}\BibitemShut {NoStop}%
\bibitem [{\citenamefont {Lin}\ \emph {et~al.}(2013)\citenamefont {Lin},
  \citenamefont {Zhu}, \citenamefont {Fauqu\'e},\ and\ \citenamefont
  {Behnia}}]{STOEf}%
  \BibitemOpen
  \bibfield  {author} {\bibinfo {author} {\bibfnamefont {X.}~\bibnamefont
  {Lin}}, \bibinfo {author} {\bibfnamefont {Z.}~\bibnamefont {Zhu}}, \bibinfo
  {author} {\bibfnamefont {B.}~\bibnamefont {Fauqu\'e}}, \ and\ \bibinfo
  {author} {\bibfnamefont {K.}~\bibnamefont {Behnia}},\ }\href {\doibase
  10.1103/PhysRevX.3.021002} {\bibfield  {journal} {\bibinfo  {journal} {Phys.
  Rev. X}\ }\textbf {\bibinfo {volume} {3}},\ \bibinfo {pages} {021002}
  (\bibinfo {year} {2013})}\BibitemShut {NoStop}%
\bibitem [{\citenamefont {Chakraborty}\ and\ \citenamefont {Piazza}()}]{Note1}%
  \BibitemOpen
  \bibfield  {author} {\bibinfo {author} {\bibfnamefont {A.}~\bibnamefont
  {Chakraborty}}\ and\ \bibinfo {author} {\bibfnamefont {F.}~\bibnamefont
  {Piazza}},\ }\href@noop {} {\bibinfo  {journal} {to appear}\ }\BibitemShut
  {NoStop}%
\bibitem [{\citenamefont {Giuliani}\ and\ \citenamefont
  {Quinn}(1982)}]{giuliani1982lifetime}%
  \BibitemOpen
\bibfield  {journal} {  }\bibfield  {author} {\bibinfo {author} {\bibfnamefont
  {G.~F.}\ \bibnamefont {Giuliani}}\ and\ \bibinfo {author} {\bibfnamefont
  {J.~J.}\ \bibnamefont {Quinn}},\ }\href@noop {} {\bibfield  {journal}
  {\bibinfo  {journal} {Physical Review B}\ }\textbf {\bibinfo {volume} {26}},\
  \bibinfo {pages} {4421} (\bibinfo {year} {1982})}\BibitemShut {NoStop}%
\bibitem [{\citenamefont {Zheng}\ and\ \citenamefont
  {Das~Sarma}(1996)}]{SSarma}%
  \BibitemOpen
  \bibfield  {author} {\bibinfo {author} {\bibfnamefont {L.}~\bibnamefont
  {Zheng}}\ and\ \bibinfo {author} {\bibfnamefont {S.}~\bibnamefont
  {Das~Sarma}},\ }\href {\doibase 10.1103/PhysRevB.53.9964} {\bibfield
  {journal} {\bibinfo  {journal} {Phys. Rev. B}\ }\textbf {\bibinfo {volume}
  {53}},\ \bibinfo {pages} {9964} (\bibinfo {year} {1996})}\BibitemShut
  {NoStop}%
\bibitem [{\citenamefont {Jungwirth}\ and\ \citenamefont
  {MacDonald}(1996)}]{RPASelf}%
  \BibitemOpen
  \bibfield  {author} {\bibinfo {author} {\bibfnamefont {T.}~\bibnamefont
  {Jungwirth}}\ and\ \bibinfo {author} {\bibfnamefont {A.~H.}\ \bibnamefont
  {MacDonald}},\ }\href {\doibase 10.1103/PhysRevB.53.7403} {\bibfield
  {journal} {\bibinfo  {journal} {Phys. Rev. B}\ }\textbf {\bibinfo {volume}
  {53}},\ \bibinfo {pages} {7403} (\bibinfo {year} {1996})}\BibitemShut
  {NoStop}%
\bibitem [{\citenamefont {Torre}\ \emph {et~al.}(2013)\citenamefont {Torre},
  \citenamefont {Diehl}, \citenamefont {Lukin}, \citenamefont {Sachdev},\ and\
  \citenamefont {Strack}}]{Emanuele2013}%
  \BibitemOpen
  \bibfield  {author} {\bibinfo {author} {\bibfnamefont {E.~G.~D.}\
  \bibnamefont {Torre}}, \bibinfo {author} {\bibfnamefont {S.}~\bibnamefont
  {Diehl}}, \bibinfo {author} {\bibfnamefont {M.~D.}\ \bibnamefont {Lukin}},
  \bibinfo {author} {\bibfnamefont {S.}~\bibnamefont {Sachdev}}, \ and\
  \bibinfo {author} {\bibfnamefont {P.}~\bibnamefont {Strack}},\ }\href
  {\doibase 10.1103/PhysRevA.87.023831} {\bibfield  {journal} {\bibinfo
  {journal} {Phys. Rev. A}\ }\textbf {\bibinfo {volume} {87}},\ \bibinfo
  {pages} {023831} (\bibinfo {year} {2013})}\BibitemShut {NoStop}%
\bibitem [{\citenamefont {Sieberer}\ \emph {et~al.}(2016)\citenamefont
  {Sieberer}, \citenamefont {Buchhold},\ and\ \citenamefont
  {Diehl}}]{Sieberer_2016}%
  \BibitemOpen
  \bibfield  {author} {\bibinfo {author} {\bibfnamefont {L.~M.}\ \bibnamefont
  {Sieberer}}, \bibinfo {author} {\bibfnamefont {M.}~\bibnamefont {Buchhold}},
  \ and\ \bibinfo {author} {\bibfnamefont {S.}~\bibnamefont {Diehl}},\ }\href
  {\doibase 10.1088/0034-4885/79/9/096001} {\bibfield  {journal} {\bibinfo
  {journal} {Reports on Progress in Physics}\ }\textbf {\bibinfo {volume}
  {79}},\ \bibinfo {pages} {096001} (\bibinfo {year} {2016})}\BibitemShut
  {NoStop}%
\bibitem [{\citenamefont {Lang}\ \emph {et~al.}(2018)\citenamefont {Lang},
  \citenamefont {Chang},\ and\ \citenamefont
  {Piazza}}]{lang2018nonequilibrium}%
  \BibitemOpen
  \bibfield  {author} {\bibinfo {author} {\bibfnamefont {J.}~\bibnamefont
  {Lang}}, \bibinfo {author} {\bibfnamefont {D.~E.}\ \bibnamefont {Chang}}, \
  and\ \bibinfo {author} {\bibfnamefont {F.}~\bibnamefont {Piazza}},\
  }\href@noop {} {\enquote {\bibinfo {title} {Non-equilibrium diagrammatic
  approach to strongly interacting photons},}\ } (\bibinfo {year} {2018}),\
  \Eprint {http://arxiv.org/abs/1810.12921} {arXiv:1810.12921
  [cond-mat.quant-gas]} \BibitemShut {NoStop}%
\bibitem [{\citenamefont {Klapwijk}\ and\ \citenamefont
  {de~Visser}(2020)}]{klapwijk2020discovery}%
  \BibitemOpen
  \bibfield  {author} {\bibinfo {author} {\bibfnamefont {T.}~\bibnamefont
  {Klapwijk}}\ and\ \bibinfo {author} {\bibfnamefont {P.}~\bibnamefont
  {de~Visser}},\ }\href@noop {} {\bibfield  {journal} {\bibinfo  {journal}
  {Annals of Physics}\ }\textbf {\bibinfo {volume} {417}},\ \bibinfo {pages}
  {168104} (\bibinfo {year} {2020})}\BibitemShut {NoStop}%
\bibitem [{\citenamefont {Tikhonov}\ \emph {et~al.}(2020)\citenamefont
  {Tikhonov}, \citenamefont {Semenov}, \citenamefont {Devyatov},\ and\
  \citenamefont {Skvortsov}}]{tikhonov2020microwave}%
  \BibitemOpen
  \bibfield  {author} {\bibinfo {author} {\bibfnamefont {K.~S.}\ \bibnamefont
  {Tikhonov}}, \bibinfo {author} {\bibfnamefont {A.~V.}\ \bibnamefont
  {Semenov}}, \bibinfo {author} {\bibfnamefont {I.~A.}\ \bibnamefont
  {Devyatov}}, \ and\ \bibinfo {author} {\bibfnamefont {M.~A.}\ \bibnamefont
  {Skvortsov}},\ }\href@noop {} {\bibfield  {journal} {\bibinfo  {journal}
  {Annals of Physics}\ }\textbf {\bibinfo {volume} {417}},\ \bibinfo {pages}
  {168101} (\bibinfo {year} {2020})}\BibitemShut {NoStop}%
\bibitem [{\citenamefont {Beyl}\ \emph {et~al.}(2018)\citenamefont {Beyl},
  \citenamefont {Goth},\ and\ \citenamefont {Assaad}}]{QMC}%
  \BibitemOpen
  \bibfield  {author} {\bibinfo {author} {\bibfnamefont {S.}~\bibnamefont
  {Beyl}}, \bibinfo {author} {\bibfnamefont {F.}~\bibnamefont {Goth}}, \ and\
  \bibinfo {author} {\bibfnamefont {F.~F.}\ \bibnamefont {Assaad}},\ }\href
  {\doibase 10.1103/PhysRevB.97.085144} {\bibfield  {journal} {\bibinfo
  {journal} {Phys. Rev. B}\ }\textbf {\bibinfo {volume} {97}},\ \bibinfo
  {pages} {085144} (\bibinfo {year} {2018})}\BibitemShut {NoStop}%
\bibitem [{\citenamefont {Halati}\ \emph {et~al.}(2020)\citenamefont {Halati},
  \citenamefont {Sheikhan},\ and\ \citenamefont {Kollath}}]{MPS}%
  \BibitemOpen
  \bibfield  {author} {\bibinfo {author} {\bibfnamefont {C.-M.}\ \bibnamefont
  {Halati}}, \bibinfo {author} {\bibfnamefont {A.}~\bibnamefont {Sheikhan}}, \
  and\ \bibinfo {author} {\bibfnamefont {C.}~\bibnamefont {Kollath}},\ }\href
  {\doibase 10.1103/PhysRevResearch.2.043255} {\bibfield  {journal} {\bibinfo
  {journal} {Phys. Rev. Research}\ }\textbf {\bibinfo {volume} {2}},\ \bibinfo
  {pages} {043255} (\bibinfo {year} {2020})}\BibitemShut {NoStop}%
\bibitem [{\citenamefont {de~Silva}\ \emph {et~al.}(2017)\citenamefont
  {de~Silva}, \citenamefont {Zhu},\ and\ \citenamefont {Van~Voorhis}}]{DFT}%
  \BibitemOpen
  \bibfield  {author} {\bibinfo {author} {\bibfnamefont {P.}~\bibnamefont
  {de~Silva}}, \bibinfo {author} {\bibfnamefont {T.}~\bibnamefont {Zhu}}, \
  and\ \bibinfo {author} {\bibfnamefont {T.}~\bibnamefont {Van~Voorhis}},\
  }\href {\doibase 10.1063/1.4973728} {\bibfield  {journal} {\bibinfo
  {journal} {The Journal of Chemical Physics}\ }\textbf {\bibinfo {volume}
  {146}},\ \bibinfo {pages} {024111} (\bibinfo {year} {2017})},\ \Eprint
  {http://arxiv.org/abs/https://doi.org/10.1063/1.4973728}
  {https://doi.org/10.1063/1.4973728} \BibitemShut {NoStop}%
\bibitem [{\citenamefont {Karrasch}\ \emph {et~al.}(2008)\citenamefont
  {Karrasch}, \citenamefont {Hedden}, \citenamefont {Peters}, \citenamefont
  {Pruschke}, \citenamefont {Schönhammer},\ and\ \citenamefont
  {Meden}}]{Karrasch_2008}%
  \BibitemOpen
  \bibfield  {author} {\bibinfo {author} {\bibfnamefont {C.}~\bibnamefont
  {Karrasch}}, \bibinfo {author} {\bibfnamefont {R.}~\bibnamefont {Hedden}},
  \bibinfo {author} {\bibfnamefont {R.}~\bibnamefont {Peters}}, \bibinfo
  {author} {\bibfnamefont {T.}~\bibnamefont {Pruschke}}, \bibinfo {author}
  {\bibfnamefont {K.}~\bibnamefont {Schönhammer}}, \ and\ \bibinfo {author}
  {\bibfnamefont {V.}~\bibnamefont {Meden}},\ }\href {\doibase
  10.1088/0953-8984/20/34/345205} {\bibfield  {journal} {\bibinfo  {journal}
  {Journal of Physics: Condensed Matter}\ }\textbf {\bibinfo {volume} {20}},\
  \bibinfo {pages} {345205} (\bibinfo {year} {2008})}\BibitemShut {NoStop}%
\bibitem [{\citenamefont {Negele}\ and\ \citenamefont
  {Orland}(2002)}]{Negele_Orland}%
  \BibitemOpen
  \bibfield  {author} {\bibinfo {author} {\bibfnamefont {J.~W.}\ \bibnamefont
  {Negele}}\ and\ \bibinfo {author} {\bibfnamefont {H.}~\bibnamefont
  {Orland}},\ }\href@noop {} {\emph {\bibinfo {title} {Quantum Many Body
  Systems}}}\ (\bibinfo  {publisher} {Oxford University Press on Demand},\
  \bibinfo {year} {2002})\BibitemShut {NoStop}%
\bibitem [{\citenamefont {Altland}\ and\ \citenamefont
  {Simons}(2010)}]{altland}%
  \BibitemOpen
  \bibfield  {author} {\bibinfo {author} {\bibfnamefont {A.}~\bibnamefont
  {Altland}}\ and\ \bibinfo {author} {\bibfnamefont {B.~D.}\ \bibnamefont
  {Simons}},\ }\href@noop {} {\emph {\bibinfo {title} {Condensed matter field
  theory}}}\ (\bibinfo  {publisher} {Cambridge University Press},\ \bibinfo
  {year} {2010})\BibitemShut {NoStop}%
\bibitem [{\citenamefont {Pines}\ and\ \citenamefont
  {Nozières}(2018)}]{FermiLiquid}%
  \BibitemOpen
  \bibfield  {author} {\bibinfo {author} {\bibfnamefont {D.}~\bibnamefont
  {Pines}}\ and\ \bibinfo {author} {\bibfnamefont {P.}~\bibnamefont
  {Nozières}},\ }\href {\doibase 10.1201/9780429492679} {\emph {\bibinfo
  {title} {The theory of quantum liquids normal fermi liquids}}}\ (\bibinfo
  {year} {2018})\ pp.\ \bibinfo {pages} {1--180}\BibitemShut {NoStop}%
\end{thebibliography}%


\begin{thebibliography}{13}%
\makeatletter
\providecommand \@ifxundefined [1]{%
 \@ifx{#1\undefined}
}%
\providecommand \@ifnum [1]{%
 \ifnum #1\expandafter \@firstoftwo
 \else \expandafter \@secondoftwo
 \fi
}%
\providecommand \@ifx [1]{%
 \ifx #1\expandafter \@firstoftwo
 \else \expandafter \@secondoftwo
 \fi
}%
\providecommand \natexlab [1]{#1}%
\providecommand \enquote  [1]{``#1''}%
\providecommand \bibnamefont  [1]{#1}%
\providecommand \bibfnamefont [1]{#1}%
\providecommand \citenamefont [1]{#1}%
\providecommand \href@noop [0]{\@secondoftwo}%
\providecommand \href [0]{\begingroup \@sanitize@url \@href}%
\providecommand \@href[1]{\@@startlink{#1}\@@href}%
\providecommand \@@href[1]{\endgroup#1\@@endlink}%
\providecommand \@sanitize@url [0]{\catcode `\\12\catcode `\$12\catcode
  `\&12\catcode `\#12\catcode `\^12\catcode `\_12\catcode `\%12\relax}%
\providecommand \@@startlink[1]{}%
\providecommand \@@endlink[0]{}%
\providecommand \url  [0]{\begingroup\@sanitize@url \@url }%
\providecommand \@url [1]{\endgroup\@href {#1}{\urlprefix }}%
\providecommand \urlprefix  [0]{URL }%
\providecommand \Eprint [0]{\href }%
\providecommand \doibase [0]{http://dx.doi.org/}%
\providecommand \selectlanguage [0]{\@gobble}%
\providecommand \bibinfo  [0]{\@secondoftwo}%
\providecommand \bibfield  [0]{\@secondoftwo}%
\providecommand \translation [1]{[#1]}%
\providecommand \BibitemOpen [0]{}%
\providecommand \bibitemStop [0]{}%
\providecommand \bibitemNoStop [0]{.\EOS\space}%
\providecommand \EOS [0]{\spacefactor3000\relax}%
\providecommand \BibitemShut  [1]{\csname bibitem#1\endcsname}%
\let\auto@bib@innerbib\@empty
\bibitem [{\citenamefont {Gao}\ \emph {et~al.}(2020)\citenamefont {Gao},
  \citenamefont {Schlawin}, \citenamefont {Buzzi}, \citenamefont {Cavalleri},\
  and\ \citenamefont {Jaksch}}]{Gao2020}%
  \BibitemOpen
  \bibfield  {author} {\bibinfo {author} {\bibfnamefont {H.}~\bibnamefont
  {Gao}}, \bibinfo {author} {\bibfnamefont {F.}~\bibnamefont {Schlawin}},
  \bibinfo {author} {\bibfnamefont {M.}~\bibnamefont {Buzzi}}, \bibinfo
  {author} {\bibfnamefont {A.}~\bibnamefont {Cavalleri}}, \ and\ \bibinfo
  {author} {\bibfnamefont {D.}~\bibnamefont {Jaksch}},\ }\href {\doibase
  10.1103/PhysRevLett.125.053602} {\bibfield  {journal} {\bibinfo  {journal}
  {Phys. Rev. Lett.}\ }\textbf {\bibinfo {volume} {125}},\ \bibinfo {pages}
  {053602} (\bibinfo {year} {2020})}\BibitemShut {NoStop}%
\bibitem [{\citenamefont {Sakudo}\ and\ \citenamefont {Unoki}(1971)}]{STO}%
  \BibitemOpen
  \bibfield  {author} {\bibinfo {author} {\bibfnamefont {T.}~\bibnamefont
  {Sakudo}}\ and\ \bibinfo {author} {\bibfnamefont {H.}~\bibnamefont {Unoki}},\
  }\href {\doibase 10.1103/PhysRevLett.26.851} {\bibfield  {journal} {\bibinfo
  {journal} {Phys. Rev. Lett.}\ }\textbf {\bibinfo {volume} {26}},\ \bibinfo
  {pages} {851} (\bibinfo {year} {1971})}\BibitemShut {NoStop}%
\bibitem [{\citenamefont {Lin}\ \emph {et~al.}(2013)\citenamefont {Lin},
  \citenamefont {Zhu}, \citenamefont {Fauqu\'e},\ and\ \citenamefont
  {Behnia}}]{STOEf}%
  \BibitemOpen
  \bibfield  {author} {\bibinfo {author} {\bibfnamefont {X.}~\bibnamefont
  {Lin}}, \bibinfo {author} {\bibfnamefont {Z.}~\bibnamefont {Zhu}}, \bibinfo
  {author} {\bibfnamefont {B.}~\bibnamefont {Fauqu\'e}}, \ and\ \bibinfo
  {author} {\bibfnamefont {K.}~\bibnamefont {Behnia}},\ }\href {\doibase
  10.1103/PhysRevX.3.021002} {\bibfield  {journal} {\bibinfo  {journal} {Phys.
  Rev. X}\ }\textbf {\bibinfo {volume} {3}},\ \bibinfo {pages} {021002}
  (\bibinfo {year} {2013})}\BibitemShut {NoStop}%
\bibitem [{\citenamefont {Kamenev}(2011)}]{kamenevbook}%
  \BibitemOpen
  \bibfield  {author} {\bibinfo {author} {\bibfnamefont {A.}~\bibnamefont
  {Kamenev}},\ }\href@noop {} {\emph {\bibinfo {title} {Field theory of
  non-equilibrium systems}}}\ (\bibinfo  {publisher} {Cambridge University
  Press},\ \bibinfo {year} {2011})\BibitemShut {NoStop}%
\bibitem [{\citenamefont {Negele}\ and\ \citenamefont
  {Orland}(2002)}]{Negele_Orland}%
  \BibitemOpen
  \bibfield  {author} {\bibinfo {author} {\bibfnamefont {J.~W.}\ \bibnamefont
  {Negele}}\ and\ \bibinfo {author} {\bibfnamefont {H.}~\bibnamefont
  {Orland}},\ }\href@noop {} {\emph {\bibinfo {title} {Quantum Many Body
  Systems}}}\ (\bibinfo  {publisher} {Oxford University Press on Demand},\
  \bibinfo {year} {2002})\BibitemShut {NoStop}%
\bibitem [{\citenamefont {Altland}\ and\ \citenamefont
  {Simons}(2010)}]{altland}%
  \BibitemOpen
  \bibfield  {author} {\bibinfo {author} {\bibfnamefont {A.}~\bibnamefont
  {Altland}}\ and\ \bibinfo {author} {\bibfnamefont {B.~D.}\ \bibnamefont
  {Simons}},\ }\href@noop {} {\emph {\bibinfo {title} {Condensed matter field
  theory}}}\ (\bibinfo  {publisher} {Cambridge University Press},\ \bibinfo
  {year} {2010})\BibitemShut {NoStop}%
\bibitem [{\citenamefont {Abrikosov}\ \emph {et~al.}(1975)\citenamefont
  {Abrikosov}, \citenamefont {Dzyaloshinskii}, \citenamefont {Gorkov},\ and\
  \citenamefont {Silverman}}]{abrikosov2012methods}%
  \BibitemOpen
  \bibfield  {author} {\bibinfo {author} {\bibfnamefont {A.~A.}\ \bibnamefont
  {Abrikosov}}, \bibinfo {author} {\bibfnamefont {I.}~\bibnamefont
  {Dzyaloshinskii}}, \bibinfo {author} {\bibfnamefont {L.~P.}\ \bibnamefont
  {Gorkov}}, \ and\ \bibinfo {author} {\bibfnamefont {R.~A.}\ \bibnamefont
  {Silverman}},\ }\href@noop {} {\emph {\bibinfo {title} {{Methods of quantum
  field theory in statistical physics}}}}\ (\bibinfo  {publisher} {Dover},\
  \bibinfo {address} {New York, NY},\ \bibinfo {year} {1975})\BibitemShut
  {NoStop}%
\bibitem [{\citenamefont {Mattuck}(1992)}]{ladder}%
  \BibitemOpen
  \bibfield  {author} {\bibinfo {author} {\bibfnamefont {R.~D.}\ \bibnamefont
  {Mattuck}},\ }\href@noop {} {\emph {\bibinfo {title} {Methods of Quantum
  Field Theory in Statistical Physics}}}\ (\bibinfo  {publisher} {Dover
  Publications, INC. New York},\ \bibinfo {year} {1992})\BibitemShut {NoStop}%
\bibitem [{\citenamefont {Pines}\ and\ \citenamefont
  {Nozières}(2018)}]{FermiLiquid}%
  \BibitemOpen
  \bibfield  {author} {\bibinfo {author} {\bibfnamefont {D.}~\bibnamefont
  {Pines}}\ and\ \bibinfo {author} {\bibfnamefont {P.}~\bibnamefont
  {Nozières}},\ }\href {\doibase 10.1201/9780429492679} {\emph {\bibinfo
  {title} {The theory of quantum liquids normal fermi liquids}}}\ (\bibinfo
  {year} {2018})\ pp.\ \bibinfo {pages} {1--180}\BibitemShut {NoStop}%
\bibitem [{\citenamefont {Zheng}\ and\ \citenamefont
  {Das~Sarma}(1996)}]{SSarma}%
  \BibitemOpen
  \bibfield  {author} {\bibinfo {author} {\bibfnamefont {L.}~\bibnamefont
  {Zheng}}\ and\ \bibinfo {author} {\bibfnamefont {S.}~\bibnamefont
  {Das~Sarma}},\ }\href {\doibase 10.1103/PhysRevB.53.9964} {\bibfield
  {journal} {\bibinfo  {journal} {Phys. Rev. B}\ }\textbf {\bibinfo {volume}
  {53}},\ \bibinfo {pages} {9964} (\bibinfo {year} {1996})}\BibitemShut
  {NoStop}%
\bibitem [{\citenamefont {Jungwirth}\ and\ \citenamefont
  {MacDonald}(1996)}]{RPASelf}%
  \BibitemOpen
  \bibfield  {author} {\bibinfo {author} {\bibfnamefont {T.}~\bibnamefont
  {Jungwirth}}\ and\ \bibinfo {author} {\bibfnamefont {A.~H.}\ \bibnamefont
  {MacDonald}},\ }\href {\doibase 10.1103/PhysRevB.53.7403} {\bibfield
  {journal} {\bibinfo  {journal} {Phys. Rev. B}\ }\textbf {\bibinfo {volume}
  {53}},\ \bibinfo {pages} {7403} (\bibinfo {year} {1996})}\BibitemShut
  {NoStop}%
\bibitem [{\citenamefont {Torre}\ \emph {et~al.}(2013)\citenamefont {Torre},
  \citenamefont {Diehl}, \citenamefont {Lukin}, \citenamefont {Sachdev},\ and\
  \citenamefont {Strack}}]{Emanuele2013}%
  \BibitemOpen
  \bibfield  {author} {\bibinfo {author} {\bibfnamefont {E.~G.~D.}\
  \bibnamefont {Torre}}, \bibinfo {author} {\bibfnamefont {S.}~\bibnamefont
  {Diehl}}, \bibinfo {author} {\bibfnamefont {M.~D.}\ \bibnamefont {Lukin}},
  \bibinfo {author} {\bibfnamefont {S.}~\bibnamefont {Sachdev}}, \ and\
  \bibinfo {author} {\bibfnamefont {P.}~\bibnamefont {Strack}},\ }\href
  {\doibase 10.1103/PhysRevA.87.023831} {\bibfield  {journal} {\bibinfo
  {journal} {Phys. Rev. A}\ }\textbf {\bibinfo {volume} {87}},\ \bibinfo
  {pages} {023831} (\bibinfo {year} {2013})}\BibitemShut {NoStop}%
\bibitem [{\citenamefont {Sieberer}\ \emph {et~al.}(2016)\citenamefont
  {Sieberer}, \citenamefont {Buchhold},\ and\ \citenamefont
  {Diehl}}]{Sieberer_2016}%
  \BibitemOpen
  \bibfield  {author} {\bibinfo {author} {\bibfnamefont {L.~M.}\ \bibnamefont
  {Sieberer}}, \bibinfo {author} {\bibfnamefont {M.}~\bibnamefont {Buchhold}},
  \ and\ \bibinfo {author} {\bibfnamefont {S.}~\bibnamefont {Diehl}},\ }\href
  {\doibase 10.1088/0034-4885/79/9/096001} {\bibfield  {journal} {\bibinfo
  {journal} {Reports on Progress in Physics}\ }\textbf {\bibinfo {volume}
  {79}},\ \bibinfo {pages} {096001} (\bibinfo {year} {2016})}\BibitemShut
  {NoStop}%
\end{thebibliography}%

\end{document}


\newcommand{\ac}[1]{\textcolor{blue}{\textsf{[AC: #1]}}}
\newcommand {\beq} {\begin{equation}}
\newcommand {\eeq} {\end{equation}}
\newcommand {\bqa} {\begin{eqnarray}}
\newcommand {\eqa} {\end{eqnarray}}
\newcommand {\ba} {\ensuremath{b^\dagger}}
\newcommand {\Ma} {\ensuremath{M^\dagger}}
\newcommand {\psia} {\ensuremath{\psi^\dagger}}
\newcommand {\psita} {\ensuremath{\tilde{\psi}^\dagger}}
\newcommand{\lp} {\ensuremath{{\lambda '}}}
\newcommand{\A} {\ensuremath{{\bf A}}}
\newcommand{\Q} {\ensuremath{{\bf Q}}}
\newcommand{\kk} {\ensuremath{{\bf k}}}
\newcommand{\qq} {\ensuremath{{\bf q}}}
\newcommand{\kp} {\ensuremath{{\bf k'}}}
\newcommand{\rr} {\ensuremath{{\bf r}}}
\newcommand{\rp} {\ensuremath{{\bf r'}}}
\newcommand {\ep} {\ensuremath{\epsilon}}
\newcommand{\nbr} {\ensuremath{\langle ij \rangle}}
\newcommand {\no} {\nonumber}
\newcommand{\up} {\ensuremath{\uparrow}}
\newcommand{\dn} {\ensuremath{\downarrow}}
\newcommand{\rcol} {\textcolor{blue}}

\title{Supplementary Material for ``Long-Range Photon Fluctuations Enhance Photon-Mediated Electron Pairing and Superconductivity ''}
\author{Ahana Chakraborty}\email{ahana@pks.mpg.de}
\author{Francesco Piazza}\email{piazza@pks.mpg.de}
\affiliation{Max Planck Institute for the Physics of Complex Systems, N\"othnitzer Str. 38, 01187, Dresden, Germany.}

\pacs{}
\date{\today}

\maketitle

\appendix

\section{Effective Electron-Electron Interaction Mediated by Photons} 
\label{sec:action}
In the main text, we analyze the pairing instability leading to superconducting transition in the coupled light-matter system. We consider a 2D electron system described by the Hamiltonian,
\beq
H_e=\sum \limits_{\vec{k},\sigma} \epsilon_{k} ~c^{\dagger}_{\vec{k},\sigma}c_{\vec{k},\sigma},
\label{supp:ham_elec}
\eeq
where $c^{\dagger}_{\vec{k},\sigma}$ is the creation operator of an electron of momentum, $\vec{k}$ and spin $\sigma$. The dispersion of the non-interacting electrons with effective mass, $m^{\ast}$ is given by $\epsilon_{k}=|\vec{k}|^2/2m^*-E_F$, where the electronic energy is measured from the Fermi surface, at $\vec{k}=\vec{k}_F$. The 2D electron system is coupled to a tera-Hertz optical cavity which supports a single mode of frequency $\omega_c$ and momentum $q_0\hat{x}$ with $q_0=\omega_c \sqrt{\epsilon_r/c}$.
The coupled light-matter system is placed under a vertical beam of laser of frequency $\omega_L$, which is detuned from the cavity frequency by $\delta_c=\omega_c - \omega_L$. We consider the detuning to be large enough so that the optical cavity is perfectly transparent at the frequency of the laser. In the frame rotating with the Laser frequency $\omega_L$, the cavity Hamiltonian takes the form, 
\beq
H_c=\delta_c~ b^{\dagger}b,
\label{supp:ham_photon}
\eeq
where $b^{\dagger}$ is the creation operator of cavity photons of momentum $\vec{q}_0$. The electron Hamiltonian is unaffected by the rotation of the reference frame. In presence of strong driving by the external Laser beam, the dominant light-matter interaction is induced via the two-photon diamagnetic process which leads to coupling between the electrons and photons in the density channel given by\cite{Gao2020},
\beq
H_{\rm{light-matter}} =\sum \limits_{ \vec{k},\sigma} \sum \limits_{\vec{q}=\pm q_0 \hat{x}}g_0 c^{\dagger}_{\vec{k}+\vec{q},\sigma}c_{\vec{k},\sigma} (b+b^{\dagger}).
\label{supp:H_lighmatter}
\eeq
Here, the strength of the light-matter coupling, $g_0 \propto \sqrt{I_d}\omega_c / \omega_L$ which is directly tunable by the intensity $I_d$ of the external laser beam\cite{Gao2020}. For the numerical calculations shown in the main text, we use the set of parameters for the electronic system consistent with 2D interface between lanthanum titanate (LAO) and strontium titanate (STO) in normal phase and for photons consistent with a split-ring cavity, given by \citep{Gao2020,STO,STOEf},
\bqa
E_F =  0.266~\mathrm{THz}~,~m^{\ast}=2m_e, ~\epsilon_{k_F+q_0}=1.4\times 10^{-3}~\mathrm{THz},~\omega_c=2\pi \times 0.3 \mathrm{THz}~,~\delta_c=0.1\omega_c ~,~\epsilon_r=10^4 .
\eqa

We use the real time formulation of non-equilibrium field theory \cite{kamenevbook} to analyze the superconducting transition in the coupled light-matter system. Schwinger-Keldysh (SK) field theory formalism provides an efficient tool to study driven-dissipative systems as well as systems in equilibrium. In the latter case, SK field theory is completely an equivalent formulation of the standard Matsubara field theory \citep{Negele_Orland}. In this appendix, we will derive the effective action describing the electron-electron interaction mediated by the cavity photons.

In SK field theory formalism, the evolution of a many body density matrix, $\hat{\rho}(t)=U(t,0)\hat{\rho}_0 U^\dagger(t,0)$, is represented by two path integrals involving two copies of fields, $\phi_+(r,t)$ and $\phi_-(r,t)$, where $r$ denotes any single particle index. These fields correspond to the forward and backward evolution of real time $t$, represented by $U(t,0)$ and $U^\dagger(t,0)$ respectively. These two path integrals are not independent, rather connected by the initial density matrix, $\hat{\rho}_0$ at $t=0$. Hence, to get rid of the redundancy, it is convenient to work in a rotated basis, $\phi_{cl,q}=(\phi_+\pm \phi_-)/\sqrt{2}$. In the $(cl,q)$ basis, we define two independent Green's functions of SK field theory, namely retarded (advanced) Green's functions, $G_{R(A)}(r,t;r',t')$ and the Keldysh Green's function, $G_K(r,t;r',t')$, given by,
\bqa
G_{R}(r,t;r',t')= \left \langle \phi_{cl}(r,t) \phi_{q}^*(r',t') \right \rangle~,
 G_{K}(r,t;r',t')= \left \langle \phi_{cl}(r,t) \phi_{cl}^*(r',t') \right \rangle.
\eqa
The causal structure of the Green's functions is ensured by $G_A=G_R^{\dagger}$ and $G_K=-G_K^{\dagger}$. $G_{R(A)}$ only contains information about spectrum and lifetime of the system, while $G_K$ explicitly depends on the distribution function of the system. 
In thermal equilibrium, these two Green's functions are related by the Fluctuation-Dissipation theorem (FDT) \cite{kamenevbook}, 
 \beq
 G_K(r,r';\omega)=2~\mathrm{Im}\left[ G_R(r,r';\omega) \right]~F(\omega),
 \eeq
 where, $F(\omega)=[\tanh(\omega/2T)]^{-\zeta}$ is the thermal distribution function of the system, with $\zeta=\pm 1$ for Bosons and Fermions respectively.

 \begin{figure}[t]
   \centering
  \includegraphics[width=0.5\textwidth]{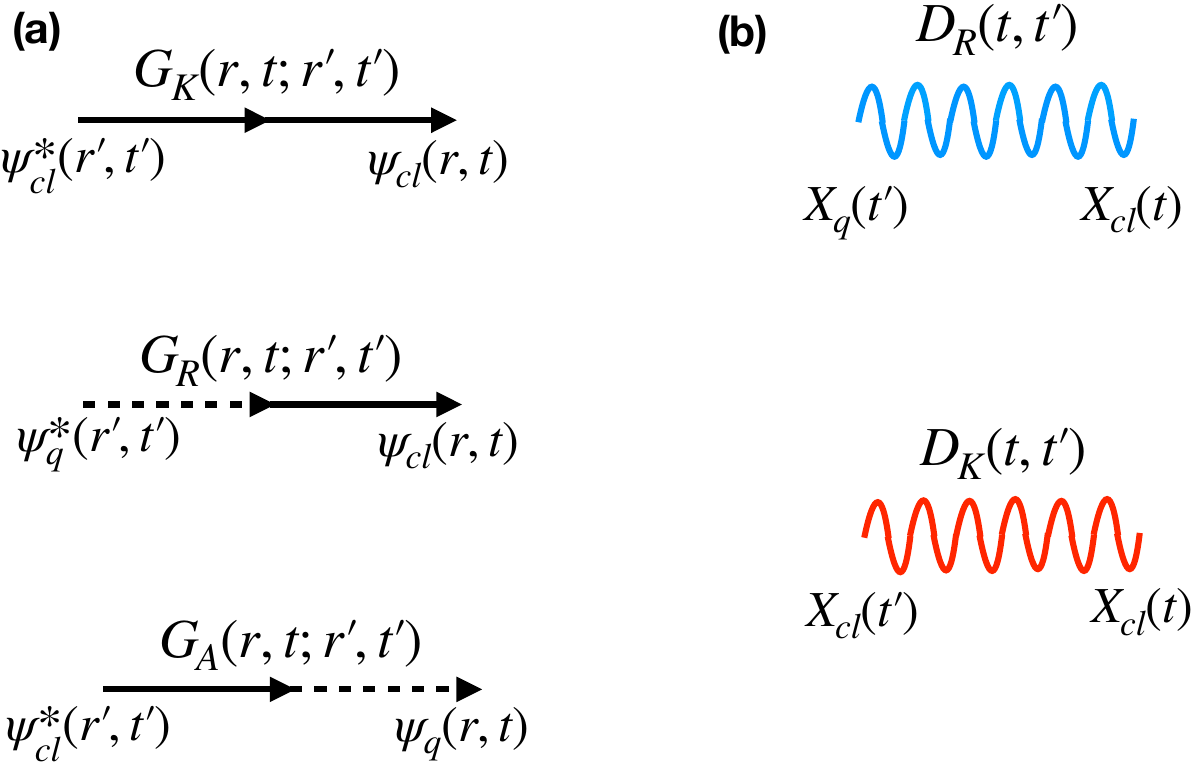}
  \caption{(a) Green's functions for Fermions, involving the Grassmann fields, $\psi_{cl,q}=(\psi_+\pm \psi_-)/\sqrt{2}$. $G_R~(G_A)$ is the retarded (advanced) component and $G_K$ is the Keldysh component. Here the classical and the quantum fields are denoted by the solid and dashed lines respectively. (b) Green's functions for the cavity photons, involving the real Bosonic fields, $X_{cl,q}=(X_+\pm X_-)/2$. The retarded component, $D_R$ is represented by the blue curvy line while the Keldysh component, $D_K$ is represented by the red curvy line.} 
   \label{fig:keldyshbasic}
   \end{figure}

In the case of our particular interest in this work, the Green's functions, $G_{R(A)}$ and $G_K$ for the 2D electrons involving the Grassmann fields $\psi_{cl(q)}(r,t)$ (coherent states of the Fermionic creation-annihilation operator) are shown in Fig. \ref{fig:keldyshbasic}(a). Here the solid and dashed lines represent the the classical and quantum Grassmann fields respectively. In this case, the Schwinger-Keldysh action for the 2D electrons described by the Hamiltonian in Eq.\ref{supp:ham_elec} is diagonal in the frequency-momentum basis and take the form,
\bqa
S_{e} &=& \int \limits_{-\infty}^{\infty} \frac{d\omega}{2\pi} \sum_{\vec{k},\sigma} \Psi^{\dagger} (\vec{k},\omega) \left[ {\begin{array}{cc}
  0 &G_A^{-1}\\ 
   G_R^{-1} & [G_K]^{-1} \\
  \end{array} } \right]\Psi (\vec{k},\omega) , \nonumber \\
  \label{supp:S_e}
\eqa
where, $\Psi^{\dagger}=[\psi_{cl,\sigma}^*~,~\psi_{q,\sigma}^*]$. Here, $G_{R(A)}^{-1}$ and $[G_{K}]^{-1}$ are the retarded (advanced) and the Keldysh component of the inverse Green's functions of electrons. For non-interacting electrons in thermal equilibrium, 
the bare inverse Green's functions are given by,
$G_{R(A)}^{-1}=w-\epsilon_{k} \pm \mathbf{i} 0^+$ and $[G_{K}]^{-1}$ is a purely imaginary regulatory term. Inverting the kernel in Eq. \ref{supp:S_e}, we obtain the thermal Green's functions of the non-interacting electrons at temperature $T$ of the form,
\bqa
G_{R(A)} (\vec{k},\omega)=\frac{1}{w-\epsilon_{k} \pm \mathbf{i} 0^+}~,~
G_{K} (\vec{k},\omega)=-2\pi \mathbf{i} \tanh\left(\frac{\omega}{2T}  \right) \delta(\omega-\epsilon_{k}).
\label{supp:Greenbare_e}
\eqa
In sections \ref{sec:BCS} and \ref{sec:Fluctuation}, we use these bare Green's functions of the thermal electrons to calculate the pairing instability of the system. These bare Green's functions will be further dressed by the self-energies induced by the screened Coulomb repulsion between the 2D electrons. This introduces finite lifetime of the quasi-particles which we will discuss in section \ref{sec:Coulomb}. 

Next, we consider the system of photons in the driven optical cavity. Guided by the form of the light-matter interaction Hamiltonian (Eq. \ref{supp:H_lighmatter}), where the density operator of the electrons couple to the position operator of the photons ${x}=(b+b^{\dagger})/\sqrt{2\delta_c}$, we represent photons by the real Bosonic fields, $X_{\pm}(t)$. They are related to the complex Bosonic fields, $\chi_{\pm}(t)$ (coherent states of the creation-annihilation operators of photons) by \cite{kamenevbook},
\beq
X_{\pm}(t) = \frac{1}{\sqrt{2 \delta_c}}\left[ \chi_{\pm}(t)+\chi^*_{\pm}(t) \right].
\eeq
The Green's functions of the photons in terms of the real scalar fields, $X_{cl(q)}=[X_+\pm X_-]/2$ are shown in Fig.\ref{fig:keldyshbasic}(b). Here, the blue and red curvy lines represent the retarded ($D_R$) and Keldysh ($D_K$) Green's functions of the photons respectively. For real Bosonic fields, $D_R(t,t')=D_A(t',t)$. In this basis, the SK action for the single mode cavity in the rotating frame takes the form,
\bqa
S_{c} &=& \int \limits_{-\infty}^{\infty} \frac{d\omega}{2\pi}  \hat{X}^{T} (\omega) \left[ {\begin{array}{cc}
  0 & D_A^{-1}\\ 
    D_R^{-1} &[ D_K]^{-1} \\
  \end{array} } \right] \hat{X}(\omega) ,~~~~
  \label{supp:S_c}
\eqa
where $\hat{X}^T=[X_{cl},X_q]$. Here,
$D_{R(A)}^{-1}$ and $[D_{K}]^{-1}$ are the retarded (advanced) and the Keldysh component of the inverse Green's functions of photons. In case of the cavity photons in thermal equilibrium at temperature $T$, 
the bare inverse Green's functions take the form,
$D_{R(A)}^{-1}=(w\pm\mathbf{i}0^+)^2-\delta_c^2$ and $[G_{K}]^{-1}$ is a purely imaginary regulatory term. Inverting the kernel in Eq. \ref{supp:S_c}, we obtain the thermal Green's functions of the bare photons at temperature $T$ of the form,
 \bqa
 \label{supp:Greenbare_p}
D_{R(A)}(\omega)= \frac{1}{2} \frac{1}{(\omega \pm \mathbf{i} 0^+)^2-\delta_c^2}~,  ~
D_K(\omega) = \frac{1}{2}~ \frac{(-\pi \mathbf{i})}{\delta_c} \left[   \delta(\omega - \delta_c) - \delta(\omega + \delta_c) \right] \coth \left( \frac{\omega}{2T} \right).
 \eqa
 We note that in case of a single mode, the dispersion of the cavity photons is replaced by the shifted energy of the mode, $\delta_c$ in the rotating frame. In the subsequent appendices \ref{sec:BCS}, \ref{sec:Fluctuation}, \ref{sec:Coulomb}, we will restrict our discussion to the case of thermal photons in equilibrium with the 2D electrons at temperature, $T$. We postpone the discussion on the case of driven-dissipative photons to appendix \ref{sec:loss}.

\begin{figure}[t]
   \centering
  \includegraphics[width=1.0\textwidth]{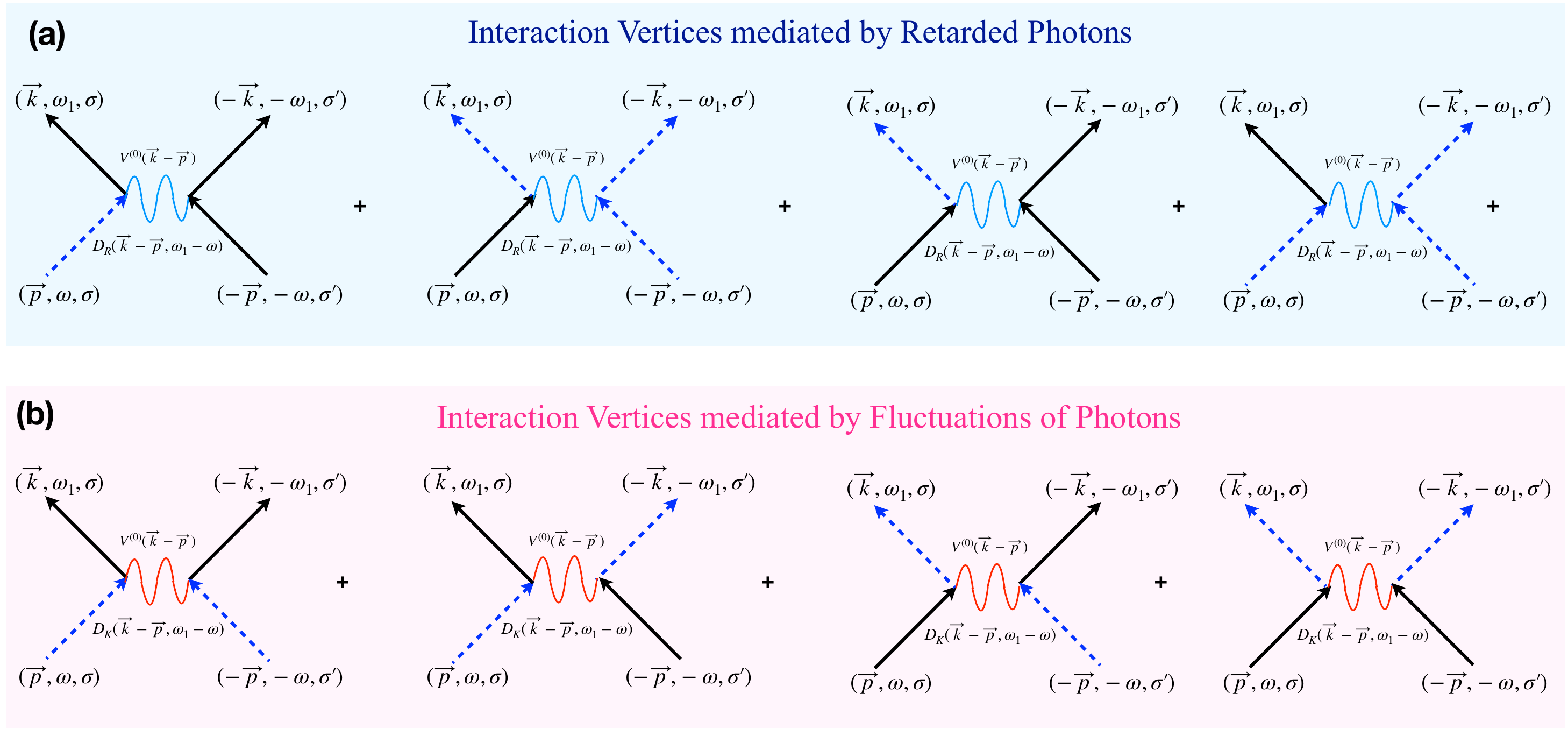}
  \caption{The light-matter coupling Hamiltonian given in Eq. [\ref{supp:H_lighmatter}] generates 12 effective electron-electron interaction vertices. Electrons and photons are represented by straight and curvy lines respectively. (a) shows the 4 interaction vertices which are mediated by the Retarded propagator of the cavity photons. Similar 4 vertices mediated by advanced propagator of photons are not shown in the figure. In the adiabatic limit, these standard ``BCS" vertices lead to attractive interactions between electrons near Fermi surface and formation of Cooper pair. (b) shows the 4 electron-electron interaction vertices which consist of Keldysh propagator of the mediator photons. These vertices explicitly depend the distribution function of the photons. We propose a new mechanism for pairing low-energy electrons, originating from these ''fluctuation" vertices, leading to significant enhancement of the pairing instability in the 2D electron system.   } 
   \label{fig:vertices}
   \end{figure}

Next, we write the Keldysh action corresponding to the light-matter interaction Hamiltonian (Eq. \ref{supp:H_lighmatter}) of the form,
\bqa
S_{\rm{light-matter}}=-g_0 \sqrt{2\delta_c} \int \limits_{-\infty}^{\infty} dt \sum \limits_{ \vec{k},\sigma} \sum \limits_{\vec{q}=\pm q_0 \hat{x}} \Big[
\{ \psi^*_{cl,\sigma}(\vec{k}&+&\vec{q},t) \psi_{q,\sigma}(\vec{k},t)+ cl \leftrightarrow q  \}X_{cl}(t) \nonumber \\
&&+
 \{ \psi^*_{cl,\sigma}(\vec{k}+\vec{q},t) \psi_{cl,\sigma}(\vec{k},t)+ cl \leftrightarrow q \}X_{q}(t)
\Big]~~~
\label{supp:S_el_ph}
\eqa
After integrating out the quadratic action of the photons, we generate the cavity-mediated effective electron-electron interaction of the form,

\bqa
&&S_{eff}=- \int \int d\omega ~d\omega_1  \sum_{\vec{p},\vec{k},\sigma,\sigma'} V^{(0)}(\vec{k}-\vec{p})\nonumber \\
&&\Bigg[ D_{R}(\omega_1-\omega) \Big[ \psi^*_{cl,\sigma} (\vec{k},\omega_1) \big\{ \psi^*_{cl,\sigma'} (-\vec{k},-\omega_1) \psi_{cl,\sigma'}(-\vec{p},-\omega) +cl \leftrightarrow q \big\} \psi_{q,\sigma} (\vec{p},\omega) \nonumber \\
&&~~~~~~~~~~~~~~~~~~~~~~~~~~~~~~~~~~~~~~~~+
 \psi^*_{q,\sigma} (\vec{k},\omega_1) \big\{ \psi^*_{cl,\sigma'} (-\vec{k},-\omega_1) \psi_{cl,\sigma'}(-\vec{p},-\omega) +cl \leftrightarrow q \big\} \psi_{cl,\sigma} (\vec{p},\omega) 
\Big] \nonumber \\
&&
+
D_{A}(\omega_1-\omega) \Big[ \psi^*_{cl,\sigma} (\vec{k},\omega_1) \big\{ \psi^*_{cl,\sigma'} (-\vec{k},-\omega_1) \psi_{q,\sigma'}(-\vec{p},-\omega) +cl \leftrightarrow q \big\} \psi_{cl,\sigma} (\vec{p},\omega) \nonumber \\
&&~~~~~~~~~~~~~~~~~~~~~~~~~~~~~~~~~~~~~~~~+
 \psi^*_{q,\sigma} (\vec{k},\omega_1) \big\{ \psi^*_{cl,\sigma'} (-\vec{k},-\omega_1) \psi_{q,\sigma'}(-\vec{p},-\omega) +cl \leftrightarrow q \big\} \psi_{q,\sigma} (\vec{p},\omega) 
\Big] 
\nonumber \\
&&
+
D_{K}(\omega_1-\omega) \Big[ \psi^*_{cl,\sigma} (\vec{k},\omega_1) \big\{ \psi^*_{cl,\sigma'} (-\vec{k},-\omega_1) \psi_{q,\sigma'}(-\vec{p},-\omega) +cl \leftrightarrow q \big\} \psi_{q,\sigma} (\vec{p},\omega) \nonumber \\
&&~~~~~~~~~~~~~~~~~~~~~~~~~~~~~~~~~~~~~~~~+
 \psi^*_{q,\sigma} (\vec{k},\omega_1) \big\{ \psi^*_{cl,\sigma'} (-\vec{k},-\omega_1) \psi_{q,\sigma'}(-\vec{p},-\omega) +cl \leftrightarrow q \big\} \psi_{cl,\sigma} (\vec{p},\omega) 
\Big] 
\Bigg].
\nonumber \\
\eqa
Here 
$
V^{(0)}(\vec{q})=g_0^2\delta_c~ \delta_{\vec{q},\pm q_0 \hat{x}}
$ is the coupling function. In the Keldysh space, this electron-electron interaction is represented by $12$ bare interaction vertices, as shown in Fig \ref{fig:vertices}. Out of them, $4$ interaction vertices are mediated by the retarded propagation of photons, $D_R$, shown by blue curvy lines in Fig \ref{fig:vertices} (a). Similar $4$ vertices are mediated by advanced propagation of photons, $D_A$ (not shown in the figure). These vertices are responsible to give rise to attractive interaction between electrons , i.e in the adiabatic limit, where the relaxation frequency, $\omega_1-\omega$, of the mediator bosons is much smaller than their characteristics energy, $\delta_c$, these vertices reduce to, 
\beq
V^{(0)}(\vec{k}-\vec{p}) D_R(\omega_1-\omega) \simeq g_0^2\delta_c ~\frac{(-1)}{2\delta_c^2} ~\delta_{\vec{k}-\vec{p},\pm q_0 \hat{x}}=-\frac{g_0^2}{2\delta_c} ~\delta_{\vec{k}-\vec{p},\pm q_0 \hat{x}}.
\label{supp:bareBCSvertex}
\eeq
This attractive interaction leads to the pairing instability to form cooper pairs via the standard BCS mechanism \citep{altland,abrikosov2012methods}.
%
%
 %

In addition to these vertices, the light-matter coupling generates $4$ more vertices which are mediated by the Keldysh propagator of the cavity photons, represented by red curvy lines in Fig \ref{fig:vertices}(b). These vertices, which are mediated by fluctuations of the photons, further assist the pairing between the low-energy electrons, leading to significant enhancement of the transition temperature, $T_c$. This is main finding of the work which will be discussed in the next sections.

We note that it has been argued in Ref [\onlinecite{Gao2020}] that at high temperatures, $T>30 mK$, this cavity mediated long ranged interaction between the two dimensional electrons can not be effectively screened by the electron gas. Hence we will restrict ourselves to this un-screened cavity-mediated e-e interaction for the subsequent analysis in this paper.

\section{Pairing Instability in Cooper Channel}
\label{sec:PairingInstability}

In this section, we will investigate the pairing instability in the 2D electron system mediated by the light-matter coupling. To obtain the superconducting transition temperature, we will analyze the divergence the four-point vertex function $\Gamma$ in the pairing channel coming from repeated scattering of the incoming electrons leading to formation of Cooper pair below the critical temperature. We denote the momenta and frequencies of the incoming electrons by $(\vec{P}+\vec{p},\Omega+\omega)$ and $(\vec{P}-\vec{p},\Omega-\omega)$ while those for the outgoing electrons are given by $(\vec{P}+\vec{p}~',\Omega+\omega')$ and $(\vec{P}-\vec{p}~',\Omega-\omega')$. In general the vertex function $\Gamma(\vec{P},\Omega;\vec{p},\omega;\vec{p}~',\omega')$ depends on three variables, which are chosen to be the center of mass (COM) coordinates, the relative incoming coordinates and the relative outgoing coordinates. 
$\Gamma$ is obtained by solving the Bethe-Salpeter (BS) equation, which is a Dyson-type equation for the four point vertex functions, calculated within the ladder approximation \cite{ladder}. This equation is in general a complicated coupled equation in relative incoming momenta and frequencies as well as in the Keldysh space between $12$ vertices shown in Fig. \ref{fig:vertices}.

\begin{figure}[t]
   \centering
  \includegraphics[width=1.0\textwidth]{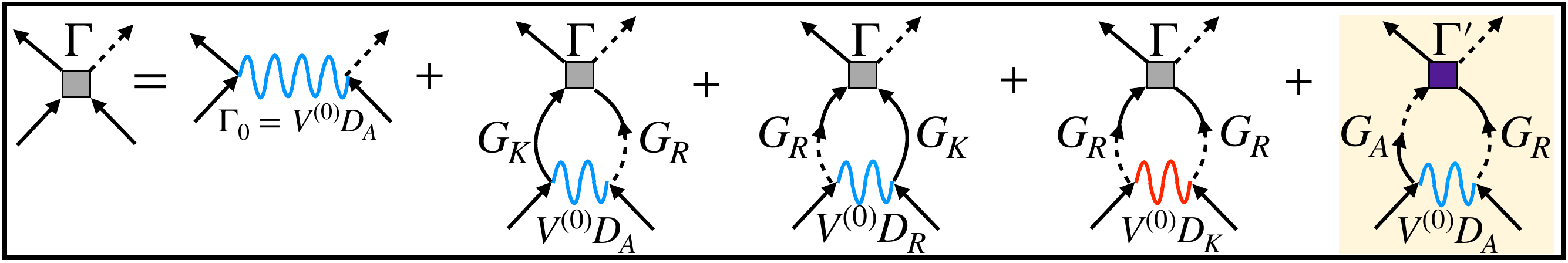}
  \caption{This Bethe-Salpeter equation for the vertex function, $\Gamma$, in the pairing channel closes in Keldysh space when of the last (highlighted) term vanishes. This is achieved in the adiabatic limit, when the mediator boson relaxes much slower than its characteristics frequency. The resultant BS equation (without the highlighted term) is a coupled equation only in incoming momenta and frequencies. The first term is the bare vertex, $\Gamma_0$. The second and third terms, $\Gamma_{\mathrm{BCS}}$, are mediated by advanced and retarded propagators of photons respectively. The fourth term, $\Gamma_{\mathrm{fluct}}$, includes the effect of photon fluctuations which leads to non-BCS type pairing mechanism described in the main text. } 
   \label{fig:BSclosed}
   \end{figure}

We will first analyze the structure of the BS equations in the Keldysh space. The  12 BS equations corresponding to 12 effective four point vertices shown in Fig. \ref{fig:vertices} have block diagonal structure in Keldysh space. The particular BS equation we will investigate to obtain $T_c$ is shown in Fig. \ref{fig:BSclosed} and given by,
\bqa
&&\Gamma(\vec{P},\Omega;\vec{p},\omega;\vec{p}~',\omega')= V^{(0)}(\vec{p}'-\vec{p}) D^{A}(\omega'-\omega) \nonumber \\
&+& \mathbf{i} \int \frac{d\vec{k}}{(2\pi)^2}  \frac{d\omega_1}{2\pi}  V^{(0)}(\vec{k}-\vec{p}) {D}_{A}(\omega_1-\omega) ~G_{K}(\vec{P}+\vec{k},\Omega+\omega_1) ~G_{R}(\vec{P}-\vec{k} ,\Omega-\omega_1) \Gamma(\vec{P},\Omega;\vec{k},\omega_1;\vec{p}~',\omega') \nonumber \\
&+&
\mathbf{i} \int \frac{d\vec{k}}{(2\pi)^2}  \frac{d\omega_1}{2\pi}  V^{(0)}(\vec{k}-\vec{p}) {D}_{R}(\omega_1-\omega) ~G_{R}(\vec{P}+\vec{k},\Omega+\omega_1) ~G_{K}(\vec{P}-\vec{k} ,\Omega-\omega_1) \Gamma(\vec{P},\Omega;\vec{k},\omega_1;\vec{p}~',\omega')\nonumber \\
&+&\mathbf{i} \int \frac{d\vec{k}}{(2\pi)^2}  \frac{d\omega_1}{2\pi}  V^{(0)}(\vec{k}-\vec{p}) {D}_{K}(\omega_1-\omega) ~G_{R}(\vec{P}+\vec{k},\Omega+\omega_1) ~G_{R}(\vec{P}-\vec{k} ,\Omega-\omega_1) \Gamma(\vec{P},\Omega;\vec{k},\omega_1;\vec{p}~',\omega') \nonumber \\
&+& \mathbf{i} \int \frac{d\vec{k}}{(2\pi)^2}  \frac{d\omega_1}{2\pi}  V^{(0)}(\vec{k}-\vec{p}) {D}_{A}(\omega_1-\omega) ~G_{A}(\vec{P}+\vec{k},\Omega+\omega_1) ~G_{R}(\vec{P}-\vec{k} ,\Omega-\omega_1) \Gamma'(\vec{P},\Omega;\vec{k},\omega_1;\vec{p}~',\omega') .
\label{BS_coupled}
\eqa
At the transition temperature, we set the COM frequency $\Omega=0$ as well as the COM momentum $\vec{P}=0$, as the Cooper pair is at rest in COM frame \cite{abrikosov2012methods}. Moreover, the above equation for $\Gamma$ is completely decoupled in the 1st and 3rd coordinates, i.e in COM coordinates and the relative coordinates of the outgoing electrons. Hence for the sake of simplicity of notation, we will denote $\Gamma$ only by keeping the relative coordinates of the incoming electrons, i.e $\Gamma(\vec{p},\omega)$ for the rest of the discussion and also in the main text.
%

 We note that in Eq. \ref{BS_coupled}, the last term in R.H.S (highlighted in Fig. \ref{fig:BSclosed}) is solely responsible for coupling the equation in Keldysh space between two different vertices $\Gamma$ and $\Gamma'$. 
By the causality structure of the retarded and advanced Green's functions, we observe that in this term all the Green's functions, $D_A(\omega_1-\omega) ,G_A(\vec{k},\omega_1)$ and $G_R(-\vec{k},-\omega_1)$ have poles w.r.t $\omega_1$ in the upper half plane in the complex frequency plane of $\omega_1$. Hence, using the contour integration method we can perform the frequency integral over $\omega_1$ by the choosing the other half plane (i.e. the lower half plane) and the above term vanishes. We note that $\Gamma(\vec{P},\Omega;\vec{k},\omega_1;\vec{p}~',\omega')$ also has pole as a function of the transferred frequency between the incoming and outgoing electrons , $\omega'-\omega_1$, in the complex plane whose residue only adds to the bare vertex part in the BS equation and will not be relevant for the discussion of finding $T_c$ here.
  Hence, in absence of the last term (highlighted) shown in Fig. \ref{fig:BSclosed}, the BS equation closes in the Keldysh space. This equation is still a coupled equation in relative incoming momentum and frequencies having the following form,
\beq
 \Gamma(\vec{p},\omega)=V^{(0)}(\vec{p}'-\vec{p}) D^{A}(\omega'-\omega)+  \Gamma_{BCS}^A(\vec{p},\omega)+ \Gamma_{BCS}^R(\vec{p},\omega)+\Gamma_{\mathrm{fluct}}(\vec{p},\omega),
 \label{supp:BS_closed}
 \eeq
 where, $\Gamma_{\rm{BCS}}=\Gamma_{\rm{BCS}}^{A}+\Gamma_{\rm{BCS}}^{R}$ (2nd and 3rd terms in RH.S of eq. \ref{BS_coupled}) are mediated by advanced and retarded propagator of the photons and $\Gamma_{\mathrm{fluct}}$ (4th term) is mediated by the fluctuations of the photons. They take the form (see Eqs. 3 of the main text),
  \begin{eqnarray}
  \label{supp:convolutions}
\Gamma_{\rm{BCS}}^{A(R)}(\vec{p},\omega)& = &\mathbf{i} \int \frac{d\vec{k}}{(2\pi)^2}  \frac{d\omega_1}{2\pi}  V^{(0)}(\vec{k}-\vec{p}) {D}_{A(R)}(\omega_1-\omega) ~G_{K(R)}(\vec{k},\omega_1) ~G_{R(K)}(-\vec{k} ,-\omega_1) \Gamma(\vec{k},\omega_1) \nonumber \\
\Gamma_{\rm{fluct}}(\vec{p},\omega) &=& \mathbf{i} \int \frac{d\vec{k}}{(2\pi)^2}  \frac{d\omega_1}{2\pi}  V^{(0)}(\vec{k}-\vec{p}) {D}_K(\omega_1-\omega) ~G_R(\vec{k},\omega_1) ~G_R(-\vec{k} ,-\omega_1) \Gamma(\vec{k},\omega_1).
\end{eqnarray}

 To solve the BS equation for finding $T_c$, we first perform the integral over the loop momentum $\vec{k}$ in Eq. \ref{supp:convolutions} straightforwardly by using the delta-function structure of $V^{(0)}$ in the transferred momentum, $\vec{k}-\vec{p}=\pm \vec{q_0}$.
Next, we do the integration over the loop frequency $\omega_1$  analytically either by using the the delta functions coming from $G_K$ and $D_K$ (see Eqs. \ref{supp:Greenbare_e} and \ref{supp:Greenbare_p}) for the bare electrons and photons (in appendices \ref{sec:BCS}, \ref{sec:Fluctuation}) or in case of broadened electrons and photons (in appendices \ref{sec:Coulomb} and \ref{sec:loss}) by using the method of contour integration in complex frequency plane by computing residues at the simple poles of the electronic or photonic Green's function. 
 This reduces the BS equation to an algebraic equation which couples the pairing amplitude $\Gamma_{\rm{on-shell}}$ of the on-shell electrons ($\omega \sim \epsilon_{k_F\pm q_0}$) to that of the off-shell electrons far off from FS ($\omega \sim \pm\delta_c,\pm 2\delta_c,..$). In the next sections, we will discuss how to close this equation only in terms of the on-shell pairing amplitude $\Gamma_{\rm{on-shell}}$ using the separation of electronic and photonic energy scales. Finally, the criterion of divergence of $\Gamma_{\rm{on-shell}}$ in this equation will give us the critical temperature of the pairing instability.

In the next sections, we will discuss the BCS mechanism and the effect of photon fluctuations on the superconducting instability. In appendices \ref{sec:BCS}, \ref{sec:Fluctuation} and \ref{sec:Coulomb}, we will restrict our discussion to the case of the closed system of 2D electrons coupled to the cavity where the total light-matter system is in thermal equilibrium. In appendices  \ref{sec:BCS} and \ref{sec:Fluctuation}, we will consider the electrons to be non-interacting while in appendix \ref{sec:Coulomb}, we will discus effect of Coulomb (screened) repulsion between them in 2D. In appendix \ref{sec:loss}, we will extend this discussion towards incorporating the effects of losses and incoherent pumping of the cavity.

  \section{BCS Pairing Mechanisms}
  \label{sec:BCS}
%

 In this section, we will evaluate the BCS terms, $\Gamma_{\rm{BCS}}^{A(R)}$ (see Eq.\ref{supp:convolutions}) using the bare thermal Green's functions of the cavity photons (Eq. \ref{supp:Greenbare_p}), the bare thermal electron Green's functions ( Eq. \ref{supp:Greenbare_e}) and the coupling function $V^{(0)}$. In the ``BCS" terms, the spectral function of thermal electrons sets the internal frequency, $\omega_1=\epsilon_{k}$, yielding,
\bqa
\Gamma_{BCS}^A(\vec{p}=p\hat{x},\omega)& = & \tilde{g} \delta_c \frac{\tanh \left( \frac{\epsilon_{p+q_0}}{2T} \right) }{\epsilon_{p+q_0}}\Bigg[ \Gamma(p+q_0,\epsilon_{p+q_0}) \left \{ 1-\left( \frac{\epsilon_{p+q_0}-\omega}{\delta_c} \right)^2 \right \}^{-1} \nonumber \\
&&~~~~~~~~~~~~~~~~~~~~~~~~~~~~~~~~~~~~~~~~~~~~~~~~~~~~~~+
\Gamma(p-q_0,\epsilon_{p-q_0}) \left \{ 1-\left( \frac{\epsilon_{p-q_0}-\omega}{\delta_c} \right)^2 \right \}^{-1} \Bigg],
\label{supp:BCSGamma}
\eqa
where, $\tilde{g} = g_0^2/(4\pi \delta_c)^2$ is the dimensionless coupling constant. Setting the external momentum, $\vec{p}\sim k_F \hat{x}$ and the frequency, $\omega=\epsilon_{k_F\pm q_0}$, we can approximate $\Gamma(k_F \pm q_0,\epsilon_{k_F \pm q_0}) $ in R.H.S by the paring amplitude of the ``on-shell" electrons for $q_0 \ll k_F$, i.e $\Gamma(k_F \pm q_0,\epsilon_{k_F \pm q_0}) \sim \Gamma_{on-shell}$. To gain a physical insight into the ``BCS" mechanism, we further simplify the above equation by taking the adiabatic limit, where the frequency of the mediator photon, $\omega_1-\omega=(\epsilon_{p\pm q_0} -\omega)<< \delta_c$. In this limit, the above equation reduces to,
%
\beq
\Gamma_{BCS}^A(k_F \hat{x},\epsilon_{k_F\pm q_0}) \sim  2 \tilde{g} \delta_c  \frac{\tanh \left( \frac{\epsilon_{p+q_0}}{2T} \right) }{\epsilon_{p+q_0}}  \Gamma_{on-shell}.
\eeq
The physical process behind the ``BCS" pairing mechanism is explained in Fig. 1(c) of the main text. In this mechanism, an adiabatic photon scatters on-shell electron pair close to FS. Below the critical temperature, $T<T_c$, these low-energy electron pair form bound state, leading to divergence of pairing amplitude of on-shell electrons, $\Gamma_{\rm{on-shell}}$. This gives an estimate of $T_c$, arising only from the standard BCS pairing mechanism, by equating the ``BCS" part of Eq. \ref{supp:BS_closed}, to $\Gamma(k_F \hat{x},\epsilon_{k_F\pm q_0})$, i.e,
\bqa
\Gamma(k_F \hat{x},\epsilon_{k_F\pm q_0})  &=&\Gamma_0+ \Gamma_{BCS}^A(k_F \hat{x},\epsilon_{k_F\pm q_0})+ \Gamma_{BCS}^R(k_F \hat{x},\epsilon_{k_F\pm q_0}) \nonumber \\
\Rightarrow \Gamma_{\rm{on-shell}} &=& \Gamma_0+ 4 \tilde{g} \delta_c  \frac{\tanh \left( \frac{\epsilon_{k_F+q_0}}{2T_c} \right) }{\epsilon_{k_F+q_0}}  \Gamma_{on-shell}.
\eqa
For divergence of the on-shell pairing amplitude $\Gamma_{\rm{on-shell}}$ we have,
\beq
 4 \tilde{g} \delta_c  \frac{\tanh \left( \frac{\epsilon_{k_F+q_0}}{2T_c^{\rm{BCS}}} \right) }{\epsilon_{k_F+q_0}} =1
\label{supp:TcBCS}
\eeq
We numerically solve this equation to obtain $T_c^{\rm{BCS}}$ which is plotted as solid line with blue circles in Fig. 2(a) of the main text. 
Further, in the limit, $\epsilon_{k_F+ \pm q_0} \ll T$, we recover the polynomial dependence of the critical temperature on the coupling strength in Ref. \onlinecite{Gao2020}, 
\beq
T_c^{\mathrm{BCS}} \sim  2 \tilde{g} \delta_c,
\label{supp:TcBCSlinear}
\eeq
for cavity-mediated long-range electron-electron interaction.

\section{Fluctuation assisted enhancement of pairing instability}
 \label{sec:Fluctuation}
  In this section, we will evaluate $\Gamma_{\rm{fluct}}$ (see Eq.\ref{supp:convolutions}) using the bare thermal Green's functions of the cavity photons (Eq. \ref{supp:Greenbare_p}), the bare thermal electron Green's functions ( Eq. \ref{supp:Greenbare_e}) and the coupling function $V^{(0)}$. $\Gamma_{\rm{fluct}}$ is mediated by the Keldysh propagator of the photons and contains information about the spectral function of photons. The density of states of the thermal cavity photons set the internal frequency $\omega_1$ close to the photon resonance frequency, i.e $\omega_1=\pm \delta_c + \omega$. Hence, in this term, the mediator photons are the non-adiabatic photons with frequency $\omega_1-\omega=\pm \delta_c$, which scatter the electrons far off the FS. This gives,
\bqa
\Gamma_{\rm{fluct}}(\vec{p}=k_F\hat{x},\omega) & \sim & -2\tilde{g} \coth \left( \frac{\delta_c}{2T} \right) \Bigg[ \Gamma(k_F\hat{x}+q_0\hat{x},\omega+\delta_c)+\Gamma(k_F\hat{x}-q_0\hat{x},\omega+\delta_c) \nonumber \\
&&~~~~~~~~~~~~~~~~~~~
+  \Gamma(k_F\hat{x}+q_0\hat{x},\omega-\delta_c)+\Gamma(k_F\hat{x}-q_0\hat{x},\omega-\delta_c) \Bigg],
\label{supp:FluctGamma}
\eqa
 where $\Gamma((k_F\pm q_0)\hat{x}, \omega \pm \delta_c)$ is the vertex function of the ``off-shell" electron pair evaluated from,  $\Gamma((k_F\pm q_0)\hat{x}, \omega \pm \delta_c)=\Gamma_0+\Gamma_{\rm{BCS}}((k_F\pm q_0)\hat{x}, \omega \pm \delta_c)+\Gamma_{\rm{fluct}}((k_F\pm q_0)\hat{x}, \omega \pm \delta_c)$ (see Eq. \ref{supp:BS_closed}). For large detuning $\delta_c$ compared to the on-shell electronic energies, the BCS contributions in the off-shell pairing amplitude at frequency $\omega\pm \delta_c$ cancel each other and the dominant contribution comes from the fluctuation term $\Gamma_{\rm{fluct}}((k_F\pm q_0)\hat{x}, \omega + \delta_c)=\Gamma_{\rm{fluct}}((k_F\pm q_0)\hat{x}, \omega - \delta_c)$.

 Now, the ``off-shell" vertex function has two contributions: (i) these ``off-shell" electrons are subsequently scattered again by a non-adiabatic photons, which brings the electron-pair back close to the FS. This mechanism is depicted in Fig. 1(d) of the main text. This gives rise to an additional pairing mechanism between ``on-shell" electrons, mediated by fluctuations of non-adiabatic photons, through virtual transition to ``off-shell" electrons at the intermediate state of the process. Substituting the ``on-shell" contribution coming from $\Gamma((k_F\pm q_0)\hat{x}, \omega \pm \delta_c)$, we get,
 \bqa
 \label{gammafluctleading}
 \Gamma_{\rm{fluct}}(\vec{p}=k_F\hat{x},\omega)&=& 16 \left(  \tilde{g} \delta_c \right)^2 \coth \left( \frac{\delta_c}{2T} \right)^2 \left[\frac{1}{\omega^2-\epsilon^2_{k_F+2q_0}} + \frac{1}{\omega^2} \right]\Gamma_{\mathrm{on-shell}}.
 \eqa
 Setting $\omega=\epsilon_{k_F+q_0}$ as before and substituting $\Gamma_{\rm{fluct}}$ in Eq. \ref{supp:BS_closed}, we obtain 

 \beq
 \Gamma_{\mathrm{on-shell}} = \Gamma_0+ \left [   4 \tilde{g} \delta_c  \frac{\tanh \left( \frac{\epsilon_{k_F+q_0}}{2T_c^{\rm{fluct}}} \right) }{\epsilon_{k_F+q_0}} +16 \left(  \tilde{g} \delta_c \right)^2 \coth \left( \frac{\delta_c}{2T_c^{\rm{fluct}}} \right)^2 \left\{\frac{1}{\epsilon_{k_F+q_0}^2-\epsilon^2_{k_F+2q_0}} + \frac{1}{\epsilon_{k_F+q_0}^2} \right \}  \right]\Gamma_{\mathrm{on-shell}}. 
 \label{supp:GammaOnshell}
 \eeq
 Here the first term within the parenthesis in R.H.S is coming from the standard BCS pairing mediated by the adiabatic photons while the second term comes from the new non-BCS pairing mediated by on-shell non-adiabatic photons.

The critical temperature, $T_c^{\rm{fluct}}$ of the 2D electrons incorporating both the BCS mechanism and fluctuation-assisted pairing by numerically solving the equation,
 \beq
 \label{supp:tcfluct_code}
 4 \tilde{g} \delta_c  \frac{\tanh \left( \frac{\epsilon_{k_F+q_0}}{2T_c^{\rm{fluct}}} \right) }{\epsilon_{k_F+q_0}} +16 \left(  \tilde{g} \delta_c \right)^2 \coth \left( \frac{\delta_c}{2T_c^{\rm{fluct}}} \right)^2 \left[\frac{1}{\epsilon_{k_F+q_0}^2-\epsilon^2_{k_F+2q_0}} + \frac{1}{\epsilon_{k_F+q_0}^2} \right] =1
 \eeq
 We plot $T_c^{\rm{fluct}}$ by green dotted line in Fig. 2(a) of the main text. We can obtain a further simplified expression of $T_c^{\rm{fluct}}$ by approximating $\coth \left( \delta_c/(2T) \right) \sim 1$ at transition temperatures in the low Kelvin regime and $\epsilon_{k_F+2q_0}/\epsilon_{k_F+q_0} \sim 2$ using linearity of the electron dispersion close to the FS, given by,
 \beq
 T_c^{\mathrm{fluct}} \sim  2 \tilde{g} \delta_c \frac{1}{1-\frac{32}{3}(\tilde{g}\delta_c)^2 \frac{1}{\epsilon_{k_F+q_0}^2}}.
 \label{supp:Tcdivergent}
 \eeq
 This clearly shows that the fluctuation-assisted pairing term (second term of Eq. \ref{supp:tcfluct_code})  modifies the corresponding BCS answer ( Eq. \ref{supp:TcBCSlinear}) by decreasing the denominator significantly from 1. For realistic parameters of 2D materials (LAO/STO) coupled to a split-ring cavity \cite{Gao2020,STO,STOEf}, the ratio
$\delta_c/\epsilon_{k_F+q_0} \sim 135$, and hence the
the denominator is reduced from
$1$ already at moderate coupling strengths $\tilde{g}\sim 0.002$. Hence the additional pairing mechanism mediated by fluctuations of photons enhances the superconducting instability between the low-energy electrons, increasing $T_c$ by order of magnitude.

 (ii) Apart from the ``on-shell" contribution of $\Gamma(k_F\pm q_0,
 \omega \pm \delta_c)$ discussed above, it has another contribution
 which connects to electrons farther off from FS at frequency
 $\omega\pm 2 \delta_c$. These off-shell electrons can be scattered back close to FS via two types of higher order scattering processes between on-shell and off-shell energy sector: (a) through $\Gamma_{\rm{BCS}}$ involving off-shell photons and (b) through $\Gamma_{\rm{fluct}}$ involving on-shell photons. The former process corrects Eq. \ref{gammafluctleading} by powers of $\sim \tilde{g}\epsilon_{k_F\pm q_0}/\delta_c$, while the latter corrects it by powers of $\tilde{g}^2$. These sub-leading corrections are suppressed as long as $\tilde{g},\epsilon_{k_F \pm q_0}/\delta_c \ll 1$ and neglected for the present discussion.
 

 At this point it is worthwhile to note that the coefficient of the fluctuation-assisted pairing term ($\propto 1/\epsilon^2_{k_F\pm q_0}$) is not regulated close to the FS and diverges in the limit $q_0\rightarrow 0$. This is an artifact of using non-interacting (bare) electron Green's functions, without considering finite lifetime of the quasi-particles near FS.
 In the next section, we will take the Coulomb interaction between electrons in the 2D material into account. We will investigate the stabilization of enhancement mechanism by incorporating the fact that quasi-particles near the FS, which are taking part in the pairing mechanism,  have only finite lifetime, dictated by Fermi-liquid nature of the 2D electron system.

\section{Stabilization of the fluctuation-assisted enhancement}
\label{sec:Coulomb}
In the previous section, we discussed the effect of photon fluctuations in enhancing the pairing instability between the low energy electrons which were assumed to be non-interacting. 
In this section, we will incorporate the effects of Coulomb repulsion between the electrons within the scope of Fermi liquid theory \cite{FermiLiquid} of the 2D electron system. The screened Coulomb interaction potential induces the self-energy\cite{SSarma} between the 2D electrons.
The real part of the self-energy renormalizes the  the effective mass of the quasi-particles and the weight of the quasi-particle peak in the spectral function. However, for the present discussion of the 2D  electrons, we will neglect these minor effects \cite{RPASelf}. At finite temperature, the dominant contribution is the broadening of the quasi-particle pole, coming from the imaginary part of the self-energy, leading to finite lifetime of the quasi-particles, given by \cite{SSarma}, 
\bqa
\frac{1}{\tau_{e,\rm{cou}}(T;\epsilon_k)}&=& \frac{\pi}{8}   \frac{T^2}{E_F} \log \left(  \frac{E_F}{T} \right) ,\mathrm{for~E_F>>T>>\epsilon_k ~ and}, \nonumber \\
&=& \frac{\pi}{8}  \frac{\epsilon_k^2}{E_F}  \log \left(  \frac{E_F }{\epsilon_k} \right) ,\mathrm{for~E_F>>\epsilon_k>>	T ~ and}
\label{supp:lifetimeRPA}
\eqa
Incorporating the quasi-particle lifetime, the bare Green's functions (Eq. \ref{supp:Greenbare_e}) of the on-shell electrons will be modified to, 
\bqa
&&\mathcal{G}_{R(A)} (\vec{k},\omega)=\frac{1}{w-\epsilon_{\vec{k}} \pm \mathbf{i}\tau^{-1}_{e,\rm{cou}}(T;\epsilon_k)}, \nonumber \\
& & \mathcal{G}_{K} (\vec{k},\omega)=\frac{1}{w-\epsilon_{\vec{k}} -  \mathbf{i}\tau^{-1}_{e,\rm{cou}}}~ \frac{ \left[ -2 \mathbf{i}\tau^{-1}_{e,\rm{cou}}(T;\epsilon_k) \tanh \left( \frac{\omega}{2T} \right) \right]}{w-\epsilon_{\vec{k}} +  \mathbf{i}\tau^{-1}_{e,\rm{cou}}}.
\label{supp:Greencou_e}
\eqa
 We note that the Green's functions of the on-shell electrons dressed by the quasi-particle lifetime will be denoted by $\mathcal{G}$.
On the other hand, the off-shell electrons having frequency close to the photon resonance frequency $ \delta_c \gg \tau^{-1}_{e,\rm{cou}}$ remain essentially unaffected by this correction and we will continue to use the bare Green's functions given in Eq.\ref{supp:Greenbare_e} for off-shell electrons. 
Using these Green's functions, we compute $\Gamma_{\rm{BCS}}$ and $\Gamma_{\rm{fluct}}$ and obtain $\Gamma_{\rm{on-shell}}$ of the form,

\bqa
 \Gamma_{\mathrm{on-shell}} &&= \Gamma_0+ 4 \tilde{g} \delta_c Re\left[  \frac{\tanh \left( \frac{\epsilon_{k_F+q_0}- \mathbf{i}\tau^{-1}_{e,\rm{cou}}(T_c;\epsilon_{k_F+q_0})}{2T_c} \right) }{\epsilon_{k_F+q_0}- \mathbf{i}\tau^{-1}_{e,\rm{cou}}(T_c;\epsilon_{k_F+q_0})} \right]  \Gamma_{\mathrm{on-shell}}+
 \nonumber \\
 &&16 \! \left(  \tilde{g} \delta_c \right)^2 \coth \left( \frac{\delta_c}{2T_c} \right)^2 Re\left[\frac{1}{\epsilon_{k_F+q_0}^2-\{\epsilon_{k_F+2q_0}-\mathbf{i}\tau^{-1}_{e,\rm{cou}}(T_c;\epsilon_{k_F+2q_0})\}^2} + \frac{1}{\epsilon_{k_F+q_0}^2+\tau^{-2}_{e,\rm{cou}}(T_c;0)} \right] \! \!  \Gamma_{\mathrm{on-shell}}~~
 \label{supp:GammaOnshell_full}
\eqa

Here, the 2nd term in R.H.S corresponds to the BCS contribution while the 3rd term corresponds to the fluctuation-assisted pairing term. Note that the finite lifetime of electrons induce leading temperature dependence in the fluctuation-assisted pairing term, while the BCS term is not affected by this correction in leading order due to smoothening of the Fermi-Dirac distribution.
A numerical estimate of $T_c$ is obtained by solving the BS equation for the divergence of the on-shell pairing amplitude from the equation,
\bqa
 &&4 \tilde{g} \delta_c Re\left[  \frac{\tanh \left( \frac{\epsilon_{k_F+q_0}- \mathbf{i}\tau^{-1}_{e,\rm{cou}}(T_c;\epsilon_{k_F+q_0})}{2T_c} \right) }{\epsilon_{k_F+q_0}- \mathbf{i}\tau^{-1}_{e,\rm{cou}}(T_c;\epsilon_{k_F+q_0})} \right]+ \nonumber \\
 &&~~~~~~~~~~~~~~~~
  16 \left(  \tilde{g} \delta_c \right)^2 \coth \left( \frac{\delta_c}{2T_c} \right)^2 Re\left[\frac{1}{\epsilon_{k_F+q_0}^2-\{\epsilon_{k_F+2q_0}-\mathbf{i}\tau^{-1}_{e,\rm{cou}}(T_c;\epsilon_{k_F+2q_0})\}^2} + \frac{1}{\epsilon_{k_F+q_0}^2+\tau^{-2}_{e,\rm{cou}}(T_c;0)} \right] =1.
\eqa

We plot this numerical answer of $T_c$ as a solid line (magenta) in Fig. 2(a) in the main text. At low temperatures $\tau_{e,\rm{cou}}^{-1}(T;\epsilon_k) \ll \epsilon_{k}$ at $k=k_F\pm q_0,k_F\pm 2q_0$ and this answer closely matches with $T_c^{\rm{fluct}}$. This figure also clearly indicates that photon fluctuations lead to a significant enhancement of $T_c$, from the BCS answer over the entire range of $\tilde{g}$ plotted in Fig. 2(a) of the main text. In high temperature limit, the quasi-particle broadening energy scale, $\tau_{e,\rm{cou}}^{-1}$, dominates over the typical energy of ``on-shell'' electrons, i.e $\tau_{e,\rm{cou}}^{-1}(T;\epsilon_k) \gg \epsilon_k$ at $k=k_F\pm q_0,k_F\pm 2q_0$. In this limit, we obtain an estimate of the critical temperature, $T_c^{\mathrm{high}}$ by solving  the equation,
\bqa
4 \tilde{g} \delta_c Re\left[  \frac{\tanh \left( \frac{- \mathbf{i}\tau^{-1}_{e,\rm{cou}}(T_c^{\rm{high}};0)}{2T_c^{\rm{high}}} \right) }{- \mathbf{i}\tau^{-1}_{e,\rm{cou}}(T_c^{\rm{high}};0)} \right]+ 
  32 \left(  \tilde{g} \delta_c \right)^2 \coth \left( \frac{\delta_c}{2T_c^{\rm{high}}} \right)^2  \frac{1}{\tau^{-2}_{e,\rm{cou}}(T_c^{\rm{high}};0)} =1,
\eqa
A simplified form the above equation, obtained by smoothening the Fermi-Dirac distribution in the BCS term, is given in Eq. 8 ~of the main text.
In the high temperature limit, $T_c^{\mathrm{high}}$ (dashed line in Fig. 2(a)) shows good agreement with the full numerical answer.

To this end, it is instructive to note that the effective interaction between the on-shell (low-energy) electrons is attractive if the effective coupling function, $\Gamma_{\mathrm{on-shell}}$ between the on-shell electrons is negative. Since in Eq. \ref{supp:GammaOnshell_full} $\Gamma_0 <0$ and the coefficient of the BCS term is positive, the criterion $\Gamma_{\mathrm{on-shell}} <0$  will be trivially ensured when the coefficient of the non-BCS pairing is also positive. In this case, the new non-BCS mechanism will contribute favorably to the attractive pairing. 
At low $T$ ($\tau_{e,\rm{cou}}^{-1}(T;\epsilon_k) \ll \epsilon_{k}$), Eq. \ref{supp:GammaOnshell_full} reduces to Eq. \ref{supp:GammaOnshell} which imposes the condition, $\epsilon_{k_F+2q_0}>\sqrt{2} \omega$ for the effective interaction to be attractive. For our particular case, $\omega=\epsilon_{k_F+q_0}$. Assuming a linear dispersion close to FS, we get $\epsilon_{k_F+2q_0}=2 \epsilon_{k_F+q_0}$ and hence the above criterion is satisfied. 
On the other hand, at large $T$ ($\tau_{e,\rm{cou}}^{-1}(T;\epsilon_k) \gg \epsilon_k$), the coefficient of the non-BCS pairing, governed only by the broadening of the quasi-particles, is always positive and hence contribute favorably to the attractive pairing.


\section{Role of cavity loss and incoherent pump}
\label{sec:loss}
In the previous sections, we considered the coupled light-matter system to be a closed system which is in thermal equilibrium at temperature, $T$. In this section , we will extend this analysis to the case of an open system with a driven-dissipative cavity. We will investigate the effects of the leakage of photons from the cavity as well as incoherent pumping of photons into the cavity. In presence of single-particle loss of photons at a rate $\gamma_{\rm{loss}}$ and incoherent single-particle pump of photons at a rate $\gamma_{\rm{pump}}$, the dynamics of the cavity photons is described by the non-equilibrium Markovian master equation for the density matrix $\hat{\rho}(t)$ \citep{Emanuele2013,Sieberer_2016},
\beq
\partial_t\hat{\rho}=-\mathbf{i}\left[H_c,\hat{\rho} \right]+\gamma_{\rm{pump}} \left(b^{\dagger} \hat{\rho} b-\frac{1}{2}\{bb^{\dagger},\hat{\rho} \}\right)+\gamma_{\rm{loss}} \left(b\hat{\rho} b^{\dagger} -\frac{1}{2}\{b^{\dagger} b,\hat{\rho} \}\right),
\eeq
where the first term in R.H.S generates the coherent dynamics by the cavity Hamiltonian $H_c$ (see eq.\ref{supp:ham_photon}). The validity of the Markovian description lies in the fact the cavity photons are driven by the external laser of frequency $\omega_L$ which is much larger than other frequency scales, i.e $\omega_L\gg\delta_c\gg \epsilon_{k_F+q_0}$. Hence, the dynamics in the rotating frame can be described very well by neglecting the memory effects in the dissipative and noise kernel induced by the external vacuum field. We convert the above master equation into an equivalent description in terms of Keldysh path integral\cite{Sieberer_2016} and the bare Green's functions of the cavity photons (see eq.\ref{supp:Greenbare_p}) are modified to,
\bqa
\label{supp:Greenloss_p}
D_{R(A)}(\omega)= \frac{1}{2} \frac{1}{(\omega \pm \mathbf{i} \kappa^+)^2-\delta_c^2}~,~
D_K( \omega) =- \mathbf{i}\frac{\gamma}{ \delta_c} \frac{\omega^2+\kappa^2+\delta_c^2}{(\omega^2-\kappa^2-\delta_c^2)^2+4\kappa^2\omega^2}.
\eqa
Here, $\kappa=(\gamma_{\rm loss}-\gamma_{\rm pump})/2$ is the net decay rate (inverse lifetime) of the cavity photons and $\gamma=(\gamma_{\rm
  loss}+\gamma_{\rm pump})/2>\kappa$ defines the total noise level. In this case, the inelastic scattering between the electrons and the cavity photons further reduces the lifetime of the quasi-particles (in addition to the Coulomb lifetime) which we take into account by calculating the Fock-diagrams\cite{kamenevbook} for the cavity induced self-energies,
\bqa
\Sigma_{R,\rm{cav}}(\omega,\vec{p})&=&\frac{\mathbf{i}}{2}\int \frac{d\vec{k}}{(2\pi)^2}\int \frac{d\omega_1}{2\pi}V^{(0)}(\vec{p}-\vec{k}) \left[D_{R}(\omega-\omega_1)G_K(\omega_1,\vec{k})+D_{K}(\omega-\omega_1)G_R(\omega_1,\vec{k})  \right]\nonumber \\
\Sigma_{K,\rm{cav}}(\omega,\vec{p})&=&\frac{\mathbf{i}}{2}\int \frac{d\vec{k}}{(2\pi)^2}\int \frac{d\omega_1}{2\pi}V^{(0)}(\vec{p}-\vec{k}) \Big[D_{R}(\omega-\omega_1)G_R(\omega_1,\vec{k})+D_{A}(\omega-\omega_1)G_A(\omega_1,\vec{k})\nonumber \\
&&~~~~~~~~~~~~~~~~~~~~~~~~~~~~~~~~~~~~~~~~~~~~~~~~~~~~~~~~~~~~~~~~~~~~~~~~~~~~~~~~~~~~~~~~~~~~~~~~~~~~~~~~+D_{K}(\omega-\omega_1)G_K(\omega_1,\vec{k})  \Big]
\eqa
We use the bare electron Green's functions given in eq. \ref{supp:Greenbare_e} and the driven-dissipative photonic Green's functions given in eq. \ref{supp:Greenloss_p} to compute the self-energies. In the leading order approximation of $\delta_c\gg \epsilon_{k_F\pm q_0}$, $\rm{Im}[\Sigma_{R,\rm{cav}}(\omega,\vec{p})]\simeq -2\tilde{g}\gamma/(1+\kappa^2/\delta_c^2)$ and $\Sigma_{K,\rm{cav}}(\omega,\vec{p})\simeq -2\mathbf{i}\tilde{g}\gamma/(1+\kappa^2/\delta_c^2)\{\tanh(\epsilon_{p+q_0}/(2T))+\tanh(\epsilon_{p-q_0}/(2T))\}$. This modifies the total inverse lifetime of the quasi-particles as,
$\tau_e^{-1}=\tau_{e, \rm cou}^{-1}+\tau_{e,\rm cav}^{-1}$,
with $\tau_{e,\rm cav}^{-1}\simeq 2\tilde{g}\gamma/(1+\kappa^2/\delta_c^2) $. The Green's functions of the on-shell electrons (see eq. \ref{supp:Greencou_e}) are modified as,
\bqa
&&\mathcal{G}_{R(A)} (\vec{k},\omega)=\frac{1}{w-\epsilon_{\vec{k}} \pm \mathbf{i}\tau^{-1}_{e}(T;\epsilon_k)}, \nonumber \\
& & \mathcal{G}_{K} (\vec{k},\omega)=\frac{1}{w-\epsilon_{\vec{k}} -  \mathbf{i}\tau^{-1}_{e}}~ \frac{ \left[ -2 \mathbf{i}\tau^{-1}_{e}(T;\epsilon_k) \tanh \left( \frac{\omega}{2T} \right) \right]}{w-\epsilon_{\vec{k}} +  \mathbf{i}\tau^{-1}_{e}},
\label{supp:Greenloss_e}
\eqa
 while the Green's functions for the off-shell electrons essentially remain unaffected and are given by the bare electronic Green's functions in eq. \ref{supp:Greenbare_e}.
 
 We will now calculate the BCS and the fluctuation (noise) terms of the BS equation (Eqs. \ref{supp:convolutions}) using the dissipative photonic Green's functions (\ref{supp:Greenloss_p}), dressed Green's functions for the on-shell electrons (\ref{supp:Greenloss_e}) and the bare Green's functions for the off-shell electrons (\ref{supp:Greenbare_e}). This modifies the BCS contribution $\Gamma_{\rm{BCS}}$ given in eq. \ref{supp:BCSGamma} to,
\bqa
\Gamma_{BCS}^{A(R)}(\vec{p}=p\hat{x},\omega)& = & \tilde{g} \delta_c \frac{\tanh \left( \frac{\epsilon_{p+q_0}}{2T} \right) }{\epsilon_{p+q_0}}\Bigg[ \left \{ 1-\left( \frac{\epsilon_{p+q_0}\mp \omega-\mathbf{i}(\kappa+\tau_e^{-1})}{\delta_c} \right)^2 \right \}^{-1} \nonumber \\
&&~~~~~~~~~~~~~~~~~~~~~~~~~~~~~~~~~~~~~~~~~~~~~~~~~~~~~~+
 \left \{ 1-\left( \frac{\epsilon_{p-q_0}\mp \omega-\mathbf{i}(\kappa+\tau_e^{-1})}{\delta_c} \right)^2 \right \}^{-1} \Bigg]~\Gamma_{\rm{on-shell}},
\label{supp:BCSGamma_loss}
\eqa
 Setting $p=k_F$ and $\omega=\epsilon_{k_F\pm q_0}$ as before, we solve the BS equation taking contributions only from the BCS terms, i.e. $\Gamma_{\rm{on-shell}}=\Gamma_0+[\Gamma_{BCS}^{A}(k_F\hat{x},\epsilon_{k_F+q_0})+\Gamma_{BCS}^{A}(k_F\hat{x},\epsilon_{k_F-q_0})+\Gamma_{BCS}^{R}(k_F\hat{x},\epsilon_{k_F+q_0})+\Gamma_{BCS}^{R}(k_F\hat{x},\epsilon_{k_F-q_0})]/2$ for a divergent on-shell paring amplitude. This gives the numerical estimate of $T_c^{\rm{BCS}}$ plotted in Fig. 2(b) of the main text by solid line with circles in the case of driven-dissipative photons. The facts that the laser detuning $\delta_c$ is much larger compared to the on-shell electronic scales and the quasi-particles are long-lived compared to cavity photons at low temperatures ($T_c^{\rm{BCS}}$ ranging from 0.01K to 0.1K ), further simply the equation for $T_c^{\rm{BCS}}$ to,
 \beq
 T_c^{\rm{BCS}}\simeq 2\tilde{g}\delta_c\frac{1}{1+\frac{\kappa^2}{\delta_c^2}}.
 \eeq
 This shows the BCS prediction for critical temperature remains essentially unaffected by cavity losses  ($\kappa \ll \delta_c$) and independent of the incoherent pumping of photons into the cavity.
 
 We now calculate the fluctuation term $\Gamma_{\rm{fluct}}$ in the BS equation (eq. (3) of the main text). In the case of the driven-dissipative cavity, the distribution function of the cavity photons is no longer sharp delta functions at the cavity resonance frequencies ($\pm \delta_c$), rather broadened by the finite cavity lifetime. In this case, $\Gamma_{\rm{fluct}}$ has contributions coming from two terms:
 $\Gamma_{\rm{fluct}}=\Gamma_{\rm{fluct}}^{\rm{photon}}+\Gamma_{\rm{fluct}}^{\rm{electron}}$ where $\Gamma_{\rm{fluct}}^{\rm{electron}}$ is the contribution when we choose the pole of the on-shell electronic Green's functions which sets the internal frequency $\omega_1=\epsilon_k-\mathbf{i}\tau_e^{-1}$, while $\Gamma_{\rm{fluct}}^{\rm{photon}}$ is the one when we choose the poles of the photonic Green's functions which sets $\omega_1=\pm \delta_c-\mathbf{i}\kappa+\omega$ (see Eq.\ref{supp:convolutions}). This modifies eq. \ref{supp:FluctGamma} to yield the fluctuation contributions,
 \bqa
 \Gamma_{\rm{fluct}}^{\rm{photon}}(\vec{p}=k_F\hat{x},\omega) & =& \tilde{g}\delta_c^2 \frac{\gamma}{\kappa} \sum \limits_{\vec{k}=(k_F\pm q_0)\hat{x}}\Bigg[ G_R(\vec{k},\omega+\delta_c-\mathbf{i}\kappa)G_R(-\vec{k},-(\omega+\delta_c-\mathbf{i}\kappa))\Gamma(\vec{k},\omega+\delta_c-\mathbf{i}\kappa) \nonumber \\
&&~~~~~~~~~~~~~~~~~~~~~~~~~~~~~~~~~~~~~~
+  G_R(\vec{k},\omega-\delta_c-\mathbf{i}\kappa)G_R(-\vec{k},-(\omega-\delta_c-\mathbf{i}\kappa))\Gamma(\vec{k},\omega-\delta_c-\mathbf{i}\kappa)  \Bigg], \nonumber \\
\Gamma_{\rm{fluct}}^{\rm{electron}}(\vec{p}=k_F\hat{x},\omega) & =&2\mathbf{i} \tilde{g}\delta_c^2 \gamma  \sum \limits_{\vec{k}=(k_F\pm q_0)\hat{x}}\frac{\left[ \left(\epsilon_k-\omega-\mathbf{i}\tau_e^{-1} \right)^2+\kappa^2+\delta_c^2 \right] \Gamma_{\rm{on-shell}}}{\left[ \left(\epsilon_k-\omega-\mathbf{i}(\tau_e^{-1}+\kappa) \right)^2-\delta_c^2\right] \left[ \left(\epsilon_k-\omega-\mathbf{i}(\tau_e^{-1}-\kappa) \right)^2-\delta_c^2 \right] \left[ \epsilon_k-\mathbf{i}\tau_e^{-1} \right]},\nonumber \\
\label{supp:FluctGamma_loss}
 \eqa
where $\Gamma((k_F\pm q_0) \hat{x}, \omega \pm \delta_c-\mathbf{i}\kappa)$ is evaluated from,  $\Gamma((k_F\pm q_0) \hat{x}, \omega \pm \delta_c-\mathbf{i}\kappa)=\Gamma_0+\Gamma_{\rm{BCS}}((k_F\pm q_0) \hat{x}, \omega \pm \delta_c-\mathbf{i}\kappa)+\Gamma_{\rm{fluct}}((k_F\pm q_0) \hat{x}, \omega \pm \delta_c-\mathbf{i}\kappa)$ (see Eq. \ref{supp:BS_closed}). As we discussed in appendix \ref{sec:Fluctuation}, we will calculate the contribution of $\Gamma((k_F\pm q_0) \hat{x}, \omega \pm \delta_c-\mathbf{i}\kappa)$ to the on-shell pairing amplitude and substitute in $ \Gamma_{\rm{fluct}}^{\rm{photon}}$ which will close the BS equation in frequency space. We can compute each term of $\Gamma((k_F\pm q_0) \hat{x}, \omega \pm \delta_c-\mathbf{i}\kappa)$ from,
\bqa
\Gamma_{\rm{BCS}}^{A}(\vec{p}=p\hat{x},\omega\pm \delta_c-\mathbf{i}\kappa)& = &-\tilde{g} \delta_c^2 \sum \limits_{\vec{k}=\vec{p}\pm q_0\hat{x}} \Bigg[ \frac{ \pm 1}{2} \mathcal{G}_K(\vec{k},\omega)\mathcal{G}_R(-\vec{k},-\omega)\nonumber \\
&&
+\tanh\frac{\epsilon_k-\mathbf{i}\tau_e^{-1}}{2T}\frac{\delta_c}{\epsilon_k-\mathbf{i}\tau_e^{-1}}\frac{1}{\left( \epsilon_k-\omega\mp 2\delta_c-\mathbf{i}\tau_e^{-1}\right)\left( \epsilon_k-\omega-\mathbf{i}\tau_e^{-1}\right) }\Bigg]\Gamma_{\rm{on-shell}} \nonumber \\
\Gamma_{\rm{BCS}}^{R}(\vec{p}=p\hat{x},\omega\pm \delta_c-\mathbf{i}\kappa)& = &-\tilde{g} \delta_c^3 \sum \limits_{\vec{k}=\vec{p}\pm q_0\hat{x}} \Bigg[
\frac{\tanh\frac{\epsilon_k-\mathbf{i}\tau_e^{-1}}{2T}}{\epsilon_k-\mathbf{i}\tau_e^{-1}}\frac{\Gamma_{\rm{on-shell}}}{\left( \epsilon_k+\omega\pm 2\delta_c-\mathbf{i}(\tau_e^{-1}+2\kappa)\right)\left( \epsilon_k+\omega-\mathbf{i}(\tau_e^{-1}+2\kappa)\right) }\Bigg]\nonumber \\
\Gamma_{\rm{fluct}}(\vec{p}=p\hat{x},\omega\pm \delta_c-\mathbf{i}\kappa)& = &\tilde{g} \delta_c^2 \sum \limits_{\vec{k}=\vec{p}\pm q_0\hat{x}}\Bigg[\frac{\gamma}{2\kappa}  \mathcal{G}_R(\vec{k},\omega)\mathcal{G}_R(-\vec{k},-\omega)\nonumber \\
&&~~~~~~~~~~~~~~~~~~~~~~~
-4\mathbf{i}\gamma \tilde{D}_K(-\epsilon_k+\mathbf{i}\tau_e^{-1}-(\omega\pm \delta_c-\mathbf{i}\kappa)) \mathcal{G}_R(-\vec{k},-\epsilon_k+\mathbf{i}\tau_e^{-1})\Bigg]\Gamma_{\rm{on-shell}},
\eqa
 where,
 $\tilde{D}_K(\omega)=(\omega^2+\kappa^2+\delta_c^2)/[\{(\omega-\mathbf{i}\kappa)^2-\delta_c^2\}\{(\omega+\mathbf{i}\kappa)^2-\delta_c^2\}]$. Using
 the above equations and setting $p=k_F$ and $\omega=\epsilon_{k_F\pm
   q_0}$, we obtain $\Gamma_{\rm{fluct}}^{\rm{photon}}$ and hence
 $\Gamma_{\rm{fluct}}$ in terms of
 $\Gamma_{\rm{on-shell}}$. Substituting in eq.\ref{supp:BS_closed}, we
 solve the BS equation for the divergent on-shell pairing
 amplitude. This gives a numerical estimate of $T_c^{\rm{noise}}$
 which is plotted in Fig. 2(b) of the main text. Following arguments
 similar to the ones given in appendix \ref{sec:Fluctuation}, one can show that the dominant contribution in $\Gamma((k_F\pm q_0) \hat{x}, \omega \pm \delta_c-\mathbf{i}\kappa)$ will come from $\Gamma_{\rm{fluct}}((k_F\pm q_0) \hat{x}, \omega \pm \delta_c-\mathbf{i}\kappa)$ in the limit of large detuning $\delta_c$ compared to the electronic scales and $\kappa$. Moreover, for $\tilde{g}\delta_c^2/\kappa^2\gg 1$ , $ \Gamma_{\rm{fluct}}^{\rm{photon}}$ dominates over $ \Gamma_{\rm{fluct}}^{\rm{electron}}$.
In this limit, at high temperatures where the quasi-particle broadening $\tau_e^{-1} \gg \epsilon_{k_F\pm q_0}$, a simplified expression for $T_c^{\rm{noise}}$ is given in eq. (11) of the main text.

\bibliographystyle{apsrev4-1}
\bibliography{LIS.bib}